\documentclass[11pt,epsf]{article}
\usepackage[dvips]{epsfig}
\usepackage{amsmath}
\usepackage{amssymb}
\usepackage{theorem}
\usepackage{latexsym}
\usepackage{graphicx}

\newcommand{\rem}[1]{}

\newcommand{\R}{{\mathbb R}}
\newcommand{\C}{{\mathbb C}}

\title{\bf
Bifurcation analysis of a normal form for excitable media: Are stable
dynamical alternans on a ring possible?}
\author 
{
Georg A. Gottwald\\
{\em School of Mathematics \& Statistics, University of
Sydney,}\\
{\em NSW 2006, Australia.}\\
{\small gottwald@maths.usyd.edu.au}\\
{\hphantom{asd}}\\
{\rm dedicated to Lorenz Kramer;}\\ 
{\rm remembering and missing his enthusiasm and creative mind.}
}

\date{}

\begin{document}

\begin{titlepage}
\setcounter{page}{1}

\vfill
\maketitle

\begin{abstract}
\noindent  
We present a bifurcation analysis of a normal form for travelling
waves in one-dimensional excitable media. The normal form which has
been recently proposed on phenomenological grounds is given in form of
a differential delay equation. The normal form exhibits a symmetry
preserving Hopf bifurcation which may coalesce with a saddle-node in a
Bogdanov-Takens point, and a symmetry breaking spatially inhomogeneous
pitchfork bifurcation. We study here the Hopf bifurcation for the
propagation of a single pulse in a ring by means of a center manifold
reduction, and for a wave train by means of a multiscale analysis
leading to a real Ginzburg-Landau equation as the corresponding
amplitude equation. Both, the center manifold reduction and the
multiscale analysis show that the Hopf bifurcation is always
subcritical independent of the parameters. This may have links to
cardiac alternans which have so far been believed to be stable
oscillations emanating from a supercritical bifurcation. We discuss
the implications for cardiac alternans and revisit the instability in
some excitable media where the oscillations had been believed to be
stable. In particular, we show that our condition for the onset of the
Hopf bifurcation coincides with the well known restitution condition
for cardiac alternans.\\

\noindent
MCS: 37L10, 35K57\\

\end{abstract}
\vfill

\end{titlepage}


\noindent
{\bf{ Excitable media are abundant in nature. Examples range from
small scale systems such as intracellular calcium waves to large scale
systems such as cardiac tissue. There exists a plethora of models
describing excitable media, each of those particular to the
microscopic details of the underlying biological, chemical or physical
system. However, excitable media have certain features which are
common to all these systems. In a recent paper we introduced a normal
form for travelling waves in one-dimensional excitable media which
contains all bifurcations occurring in excitable media. The normal
form consists of a delay-differential equation and is applicable to
systems which are close to the saddle-node bifurcation of the
traveling wave. Although the normal form has so far only been proposed
on phenomenological grounds and has not yet been rigorously derived,
its parameters could be successfully fitted to some real excitable
media with good quantitative agreement. In this work we perform a
bifurcation analysis of the occurring Hopf bifurcation. This may have
important consequences for cardiac dynamics and the understanding of
arrhythmias, in particular of alternans. Alternans describe the
scenario in cardiac tissue whereby action potential durations are
alternating periodically between short and long periods. There is an
increased interest in alternans because they are believed to trigger
spiral wave breakup in cardiac tissue and to be a precursor to
ventricular fibrillation. So far these alternans were believed to be
stable. However within the normal form we show that the Hopf
bifurcation is actually subcritical suggesting that the resulting
oscillations may be unstable.  }}


\section{Introduction}
\label{Sec-intro}

Examples of excitable media are frequently found in biological and
chemical systems. Prominent examples are cardiac and neural tissue
\cite{WinfreeBook,Davidenko}, slime mold colonies in a starving
environment \cite{dictyostelium} and intracellular calcium waves
\cite{calcium}. There are two defining features of excitable media
which are crucial to enable effective signal transmission in
biological systems such as cardiac or neural tissue: threshold
behaviour and relaxation to a stable rest state. The threshold
behaviour assures that only for large enough stimuli a signal is
produced whereas small perturbation decay immediately. For
super-threshold perturbations a signal will decay only after a long
excursion -- called action potential in the context of cardiac
dynamics -- back to its stable rest state. This relaxation allows for
repeated stimulation which is essential for wave propagation in
cardiac and neural tissue.\\ Typical solutions in one-dimensional
excitable media are wave trains. The wavelength $L$ can range from
$L=\infty$ corresponding to an isolated pulse to a minimal value $L_c$
below which propagation fails. Besides wave trains rotating spiral
waves may form in two dimensional excitable media, and scroll waves in
three dimensional excitable media
\cite{Winfree,Winfree90,Winfree94,Margerit01,Margerit02}. In the
context of cardiac excitable media propagation failure of these
solutions is often linked to clinical situations. In particular the
break-up of spiral waves has been associated with pathological cardiac
arrhythmias \cite{Chaos}. Spiral waves may be created in cardiac
tissue when wave trains propagate through inhomogeneities of the
cardiac tissue. A reentrant spiral may move around an anatomical
obstacle or around a region of partially or totally inexcitable
tissue. Once created they drive the heart at a rate much faster than
normal sinus rhythm and cause tachycardia. If these spiral waves then
subsequently breakup into multiple drifting and meandering spirals and
disintegrate into a disorganized state, fibrillation may occur with a
possible fatal result for the patient, especially when occurring in
the ventricles. It is therefore of great interest to understand the
transition from one reentrant spiral to the disorganized collection of
complex reentrant pathways. Rather than investigating the full
two-dimensional problem of spiral wave break-up which would include
interactions of numerous wave arms, one can study some aspects of
spiral wave breakup by looking at a one-dimensional slice of a spiral
i.e. at a one-dimensional wave train
\cite{Courtemanche}. A pulse circulating around a one-dimensional ring
constitutes the simplest model for a spiral rotating around an
anatomical obstacle. Such models concentrate solely on the dynamics
close to the anatomical obstacle and ignore the influence of the
dynamics of the spiral arms.\\

\noindent
Experimentally this problem has been studied since the beginning of
the last century. In \cite{Mines} the circulation of an electrical
pulse around a ring-shaped piece of atrial cardiac tissue from a
tortoise heart was investigated as a model for reentrant
activity. Recently, the experiments in \cite{Frame} investigated the
dynamics of pulse circulation around a ring-shaped piece of myocardial
tissue from a dog heart. Remarkably, it was found that oscillations of
the circulation period of the pulse occurred leading to conduction
block and subsequent termination of reentry.\\

\noindent
This indicates that a one-dimensional oscillatory instability, named
alternans, may be the mechanism triggering spiral wave breakup
\cite{Nolasco,Courtemanche,Karma_A,Karma93,Fenton02}. This instability
occurs when the circulation time of the pulse around the ring is below
a certain specific threshold. Below this threshold alternans arise and
action potential durations are alternating periodically between short
and long periods. We make here the distinction between alternans which
are mediated by external periodic stimulations
\cite{Hastings00,Guevara02,Echebarria02,Echebarria02b,Fox02,Henry05,Vinet99,Vinet03,Vinet05}
and alternans which have a purely dynamical origin. We call the latter
ones {\em dynamical alternans}. It is these dynamical alternans with
which this paper is concerned with. We will analyze dynamical
alternans using a normal form proposed in \cite{GottKram05} for
travelling waves in one-dimensional excitable media. The stability of
dynamical alternans will be determined by a center manifold theory and
by multiple scale analysis. Typically oscillatory instabilities arise
in one-dimensional media via Hopf bifurcations. These Hopf
bifurcations may be supercritical resulting in sustained stable
oscillations or subcritical leading to a collapse of the oscillations
and possibly of the pulse solution. Alternans are widely believed to
originate via a supercritical bifurcation
\cite{Courtemanche,Karma93,Karma94,Courtemanche96}. This believe is
based on numerical simulations of certain models for cardiac
dynamics. However we note that the above mentioned experiments
\cite{Mines,Frame} show that the occurrence of oscillations leads to a
subsequent termination of the pulse, which is not suggestive of a
supercritical bifurcation.\\

\noindent
Our main result for a single pulse and for a wave train in a ring is
that in the framework of our normal form alternans arise as a
subcritical Hopf bifurcation. This result is in contrast to the common
belief that alternans are stable oscillations. We corroborate our
result by numerical simulations of a model for cardiac tissue in which
previous numerical simulations suggested the occurrence of stable
oscillations. We will see that previous numerical experiments have not
been performed sufficiently long to reveal the subcritical character
of the Hopf bifurcation. The subcritical character is in agreement
with the above mentioned experiments \cite{Mines,Frame} and may
explain why alternans often trigger spiral wave breakup and are
associated with cardiac failures.
The normal form describes excitable media in parameter regions where
the system is close to the saddle-node of the travelling wave. This is
the case for models such as the FitzHugh-Nagumo model
\cite{FHN}, and is typically the case when the Hopf bifurcation occurs
close to the saddle-node bifurcation of the isolated pulse. In
particular the normal form is valid for those classes of excitable
media (or parameter regions of a particular excitable medium) where
the activator weakly interacts with the preceding inhibitor which
exponentially decays towards the homogeneous rest state. However for
other models such as the model by Echebarria and Karma
\cite{Echebarria02} there exist parameter regions where upon
decreasing the length of the ring the pulse solution is driven away
from the solitary pulse solution and does not weakly interact with the
inhibitor. Our normal form cannot describe these oscillations and we
present numerical simulations where oscillations are indeed stable in
Section~\ref{Sec-numerics}.\\

\begin{figure}
\centerline{
\psfig{file=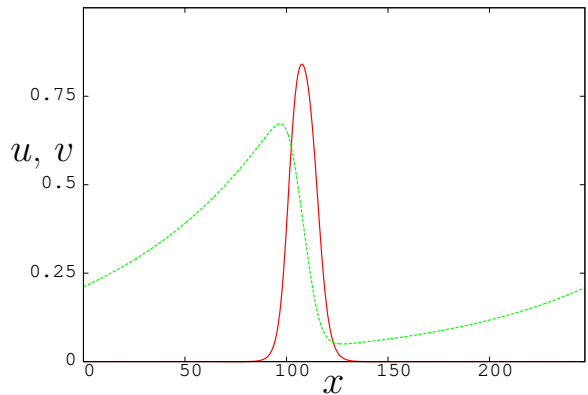,angle=0,width=3.0in}}
\caption{Plot of the activator $u$ (continuous line) and the inhibitor
$v$ (dashed line) for the modified Barkley system (\ref{barkley})
showing how the activator $u$ weakly interacts with the exponentially
decaying tail of the inhibitor $v$. The parameters are $a=0.22$,
$u_s=0.1$, $\epsilon=0.03755$. The ring length is $L=245$ which is
close to the critical value $L_c$ for the saddle-node bifurcation.}
\label{Fig-barkley2}
\end{figure}

\noindent
Recently we have constructed a normal form for travelling waves in
one-dimensional excitable media which takes the form of a delay
differential equation \cite{GottKram05} (see (\ref{NF}) and
(\ref{NF2})). The construction is based on the well-known observation
that the interaction of a pulse with the inhibitor of the preceding
pulse modifies the generic saddle-node bifurcation of an isolated
pulse. 
In Fig.~\ref{Fig-barkley2} we illustrate this scenario for a
modified Barkley model \cite{Barkley91}.
%
The normal form exhibits a rich bifurcation behaviour which we
could verify by numerically simulating partial differential equation
models of excitable media. Besides the well known saddle-node
bifurcations for isolated pulses and for periodic wave trains the
normal form also exhibits a Hopf bifurcation and a symmetry breaking,
spatially inhomogeneous pitchfork bifurcation. Moreover, the normal
form shows that the saddle-node and the Hopf bifurcation are an
unfolding of a Bogdanov-Takens point as previously suggested in
\cite{Knees92,GottwaldKramer04}. The Hopf bifurcation is found to
occur before the saddle-node bifurcation for a single pulse in a
ring. For a wave train consisting of several pulses in a ring, the
Hopf- and the saddle-node bifurcations occur after the symmetry
breaking pitchfork bifurcation in which every second pulse dies. We
could verify these scenarios in numerical simulations of a modified
Barkley-model \cite{Barkley91} and the FitzHugh-Nagumo equations
\cite{FHN}. The normal form provides a unified framework to study all
possible bifurcations of travelling waves in one-dimensional excitable
media.\\ We were able to determine the parameters of the normal form
from numerical simulations of the modified Barkley model
\cite{Barkley91,GottwaldKramer04}. Using these numerically determined
parameters we showed excellent agreement between the normal form and
the full partial differential equation. We quantitatively described
the Hopf bifurcation and the inhomogeneous pitchfork bifurcation with
the normal form. Moreover, we were able to quantify the
Bogdanov-Takens bifurcation.\\ Whereas the subcritical character of
the pitchfork bifurcation had been established in \cite{GottKram05}, a
detailed analysis of the Hopf bifurcation was missing. For example, an
interesting question is whether the Hopf bifurcation is sub-critical
(as numerically observed for the parameters chosen in
\cite{GottKram05} for the modified Barkley model), or whether it is
possible to observe sustained stable oscillations. This has important
implications for cardiac alternans as described above. In this paper
we will analyze the Hopf bifurcation of the normal form in detail. We
derive a normal form for the Hopf bifurcation which allows us to
determine the stability of the bifurcating solutions close to
criticality.\\

\noindent
Before we embark on the analytical investigation of the Hopf
bifurcation and the implications for cardiac dynamics, we state that
all conclusions drawn obviously depend on the validity of the normal
form. So far the proposed normal form which we briefly review in
Section~\ref{Sec-NF} has not been rigorously derived for any excitable
medium. However, we point out that in \cite{GottKram05} we have shown
good quantitative agreement with some real excitable media. We will
discuss the limitations of our approach in more detail in
Section~\ref{Sec-Disc}.\\

\noindent
In Section~\ref{Sec-NF} we recall the normal form and some of its
properties. In Section~\ref{Sec-BifHopf} we perform a center manifold
reduction of the normal form to describe the character of the Hopf
bifurcation for a single pulse in a ring. In Section~\ref{Sec-CGL} we
look at the case where a pseudo continuum of modes undergo a Hopf
bifurcation and derive in a multiple scale analysis a Ginzburg-Landau
equation which allows us to study the stability of the Hopf
bifurcation of a wave train. The paper concludes with
Section~\ref{Sec-Disc} where we present results from numerical
simulations, discuss the implications of our analysis to cardiac
dynamics and make connections to previous studies on alternans. In
particular, we show that our condition for the onset of the Hopf
bifurcation coincides with the well known restitution condition for
cardiac alternans.


\noindent
\section {A normal form}
\label{Sec-NF}
\vskip 5pt 

In \cite{GottKram05} we introduced a normal form for a single pulse
on a periodic domain with length $L$
\begin{eqnarray}
\label{NF}
\partial_t X = -\mu - g X^2 - \beta (\gamma + X(t-\tau) + \gamma_1X(t))\; ,
\end{eqnarray}
where
\begin{eqnarray}
\label{beta}
\beta=\beta_0 \exp{(-\kappa \tau)}\; ,
\end{eqnarray}
for positive $\beta$, $\kappa$, $\gamma$ and $\gamma_1$. Here $X(t)$
may be for example the difference of the amplitude or the velocity of
a pulse to its respective value at the saddle-node. The terms
proportional to $\beta$ incorporate finite domain effects associated
with the activator of an excitable medium running into its own
inhibitor with speed $c_0$ after the temporal delay $\tau=(L-\nu)/c_0$
where $\nu$ is the finite width of the pulse. Note that for $\beta=0$
(i.e. for the isolated pulse with $\tau \to
\infty$) we recover the generic saddle-node bifurcation which is well
known for excitable media. Numerical simulations of excitable media
show that the bifurcations of a {\it single} propagating pulse in a
ring are different from the bifurcations of a wave train consisting of
several distinct pulses. In the case of a wave train with finite wave
length where a pulse may run into the inhibitor created by its
preceding pulse, we showed that it was sufficient to consider two
alternating populations of pulses $X$ and $Y$. We derived the
following extension
\begin{eqnarray}
\label{NF2}
\partial_t X &=&-\mu - g X^2 - \beta (\gamma + Y(t-\tau) +
\gamma_1X(t)) \nonumber \\
\partial_t Y &=& -\mu - g Y^2 - \beta (\gamma + X(t-\tau) + \gamma_1Y(t))\; .
\end{eqnarray}
To avoid confusion we state that we use the term {\em normal form}
here in two different contexts. Equations (\ref{NF}) and (\ref{NF2})
are coined ``normal form'' as they are an attempt to describe the
behaviour of travelling waves in generic one-dimensional excitable
media. However, these normal forms exhibit a rather rich bifurcation
behaviour. We will focus in Section~\ref{Sec-BifHopf} on the solution
behaviour close to criticality. In that context we will speak of
``normal forms'' in the sense of bifurcation theory.


\noindent
\subsection {Properties of the normal form}
\label{Sec-BT}

\subsubsection {Saddle-node bifurcation}
\label{Sec-SN}
\begin{figure}
\centerline{
\psfig{file=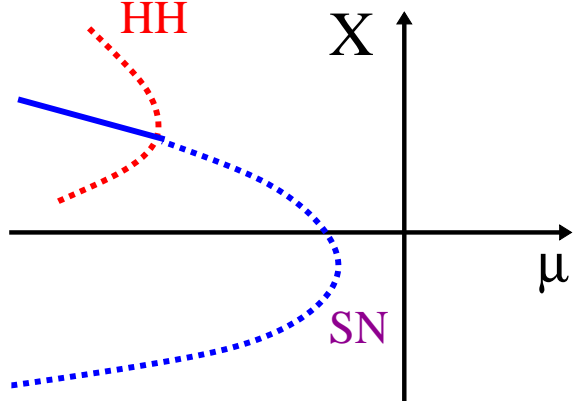,angle=0,width=3.0in}}

\caption{Sketch of the bifurcation diagram for a single
pulse in a ring showing a stationary saddle-node bifurcation (SN) and
a subcritical Hopf bifurcation (HH).}
\label{Fig-Hopfsketch}
\end{figure}

Equation (\ref{NF}) supports the following stationary solutions
\begin{eqnarray}
\label{statio}
{\bar X}_{1,2} =\frac{1}{2g}[-\beta(1+\gamma_1)\pm
\sqrt{\beta^2(1+\gamma_1)^2-4g(\mu+\beta \gamma)}]   \;.
\end{eqnarray}
It is readily seen that the upper solution branch is stable whereas
the lower one is unstable \cite{GottKram05}. The two solutions
coalesce in a saddle-node bifurcation with
\begin{eqnarray}
\label{SN_wt}
{\bar X}_{SN} = -\frac{\beta}{2g}(1+\gamma_1) \qquad {\rm{at}} \qquad
\mu_{SN}= \frac{\beta^2}{4g}(1+\gamma_1)^2-\beta \gamma \;.
\end{eqnarray}
One expects the parameter $\beta$, which describes the coupling to the
inhibitor, to be small (see \cite{GottKram05}). This implies
$\mu_{SN}< 0$, indicating that the saddle-node of a single pulse or of
a periodic wave train on a finite ring occurs at smaller values of the
bifurcation parameter $\mu$ than for the isolated pulse. Hence, the
bifurcation is shifted to the left when compared to the isolated pulse
(see Figure~\ref{Fig-Hopfsketch}).\\

\noindent
Besides this stationary saddle node bifurcation the normal form
(\ref{NF}) also contains a Hopf bifurcation.

\subsubsection{Hopf bifurcation}
\label{Sec-Hopf}
Linearization of the normal form around the homogeneous solution
${\bar X}$ with respect to small perturbations of the form $\delta X
\exp{\lambda t}$ with $\lambda=\sigma + i \omega$ yields
\begin{eqnarray}
\label{lin}
\lambda+2g{\bar X} + \beta \gamma_1 + \beta e^{-\lambda \tau} = 0\; .
\end{eqnarray}
Besides the stationary saddle-node bifurcation (\ref{SN_wt}) with
$\lambda=0$, the linearization (\ref{lin}) also reveals the existence
of a Hopf bifurcation with $\lambda=i\omega$. We readily find from
(\ref{lin})
\begin{eqnarray}
\label{HomHopfa}
\omega&=&\beta\sin{\omega \tau}\\
\label{HomHopfb}
{\bar X}_{H}&=& -\frac{\beta}{2g}(\cos{\omega \tau} + \gamma_1)\; .
\end{eqnarray}
The first equation (\ref{HomHopfa}) allows us to formulate a necessary
condition for the existence for a Hopf bifurcation
\[
\beta \tau > 1 \; ,
\]
i.e. if the coupling is strong enough and the pulse feels the presence
of the inhibitor of the preceding pulse sufficiently strongly. The
Hopf bifurcation occurs in parameter space before the saddle-node
bifurcation and bifurcates from the upper stable branch ${\bar X}_1$
of (\ref{statio}) as is readily seen by observing ${\bar X}_{H} \ge
{\bar X}_{SN}$, independent of the value of $\beta$. Hence we may
equate ${\bar X}_{H}= {\bar X}_1$ and solve for the bifurcation
parameter. Setting $\mu_{H}=\mu_{SN}-\delta\mu$ we find
\begin{eqnarray}
\label{mueHopf}
\delta \mu=\frac{\beta^2}{4g}(1-\cos{\omega \tau})^2\; .
\end{eqnarray}
%
%
%
The saddle-node bifurcation coalesces with the Hopf bifurcation in a
codimension-$2$ Bogdanov-Takens point. At the Bogdanov-Takens point
with $\mu_{H}=\mu_{SN}$ we have $\omega \tau=0$ and $\omega=0$,
i.e. the period of the oscillation goes to infinity. The amplitude at
the Bogdanov-Takens point is readily determined by comparison of
(\ref{SN_wt}) with (\ref{HomHopfb}) for $\omega \tau=0$. From
(\ref{HomHopfa}) we infer that this occurs at $\beta \tau=1$. We note
that if $\beta \tau$ is large enough there can be arbitrary many
solutions $\omega_l$ of (\ref{HomHopfa}). We will discuss this
scenario in Section~\ref{Sec-CGL}.

In Figure~\ref{Fig-Hopfsketch} we show a schematic bifurcation diagram
with the saddle-node bifurcation and the subcritical Hopf bifurcation
for a single pulse in a ring.

\subsubsection{Spatially inhomogeneous pitchfork bifurcation}
\label{Sec-PF}
When a group of several pulses in a ring is numerically simulated one
observes that this wave train group does not undergo a symmetry
preserving Hopf bifurcation on increasing the refractoriness, but
instead develops a symmetry breaking, spatially inhomogeneous
instability whereby every second pulse dies. 

The normal form (\ref{NF2}) is able to predicted and quantitatively
describe this scenario \cite{GottKram05}. The system (\ref{NF2}) for
wave trains supports two types of solutions. Besides the homogeneous
solution (\ref{statio}), ${\bar X}_h={\bar Y}_h={\bar{X}}_1$, which
may undergo a saddle-node bifurcation as described by (\ref{SN_wt}),
there exists another stationary solution, an alternating mode $X_a$
and $Y_a$, with
\begin{eqnarray}
\label{AI_2}
{\bar X}_{a}=-{\bar Y}_a+\frac{\beta}{g}(1-\gamma_1)\; .
\end{eqnarray}
Associated with this alternating solution is a pitchfork bifurcation
at
\begin{eqnarray}
\label{AI_PFa}
\mu_{PF}=\frac{1}{4}\frac{\beta^2(1+\gamma_1)^2}{g}-\frac{\beta^2}{g}-\beta
\gamma 
= \mu_{SN}-\frac{\beta^2}{g} \le \mu_{SN}\; ,
\end{eqnarray}
when
\begin{eqnarray}
\label{AI_PFb}
X_{PF}=Y_{PF}=\frac{\beta}{2g}(1-\gamma_1)\; .
\end{eqnarray}
The pitchfork bifurcation sets in before the saddle-node bifurcation
as can be readily seen from (\ref{AI_PFa}).

The upper branch of the homogeneous solution ${\bar X}_h$ given by
(\ref{statio}) at the pitchfork bifurcation point $\mu_{PF}$ coincides
with (\ref{AI_PFb}). Hence as the Hopf bifurcation, the pitchfork
bifurcation branches off the upper branch of the homogeneous
solution. The pitchfork bifurcation is always subcritical because
there are no solutions ${\bar X}_{a}$ possible for $\mu >
\mu_{PF}$. No bifurcation theory is needed to determine the
subcritical character of the pitchfork bifurcation.

The stability of the homogeneous solution ${\bar X}={\bar Y}={\bar
X}_h={\bar Y}_h$ is determined by linearization. We study
perturbations $X={\bar X}_h + x\exp{\lambda t}$ and $Y={\bar X}_h +
y\exp{\lambda t}$, and obtain as a condition for nontrivial solutions
$x$ and $y$
\begin{eqnarray}
\label{lin_ai2}
(\lambda+2g {\bar X}_h+\beta \gamma_1) = \pm \beta e^{-\lambda
\tau}\; .
\end{eqnarray}
The upper sign denotes an antisymmetric mode $x=-y$ whereas the lower
sign denotes a symmetric mode $x=y$. Stationary bifurcations are
characterized by $\lambda=0$, and in this case the symmetric mode
coincides with the saddle-node bifurcation (\ref{SN_wt}) and the
antisymmetric mode terminates at the pitchfork bifurcation
(\ref{AI_PFb}).\\

\noindent
As for the case of a single pulse in a ring, non-stationary Hopf
bifurcations are possible if $\lambda=i\omega$ for wave trains. We
obtain from (\ref{lin_ai2})
\begin{eqnarray}
\label{lin_ai3}
\omega=\mp \beta \sin{\omega \tau}
\end{eqnarray}
and
\begin{eqnarray}
\label{lin_ai4}
{\bar X}_h= \frac{\beta}{2g}(\pm\cos{\omega \tau}-\gamma_1)\; .
\end{eqnarray}
We consider only the symmetric case (the lower signs) which reproduces
our results (\ref{HomHopfa}) and (\ref{HomHopfb}) for the symmetry
preserving Hopf bifurcation. The antisymmetric case does not allow for
a single-valued positive $\omega$. For $\omega \tau\to 0$ the onset of
the Hopf bifurcation moves towards the saddle-node (\ref{SN_wt}) and
coalesces with it at $\beta \tau=1$ in a Bogdanov-Takens point as
described in Section~\ref{Sec-BT}.  For $\omega \tau\to\pi$ the Hopf
bifurcation moves towards the pitchfork bifurcation with a limiting
value of ${\bar X}_{h}=\beta(1-\gamma_1)/(2g)=X_{PF}$ in a
codimension-2 bifurcation. At the point of coalescence the Hopf
bifurcation has a period $T = 2 \tau$ which corresponds exactly to the
inhomogeneous pitchfork bifurcation with $p=\pi$ whereby every second
pulse dies. For values $\omega \tau \in [0,\pi)$ the Hopf bifurcation
always sets in after the pitchfork bifurcation. Hence for wave trains
one can see only a Hopf bifurcation of the steady solution at the
point where the Hopf bifurcation collides with the pitchfork
bifurcation.\\

\begin{figure}
\centerline{
\psfig{file=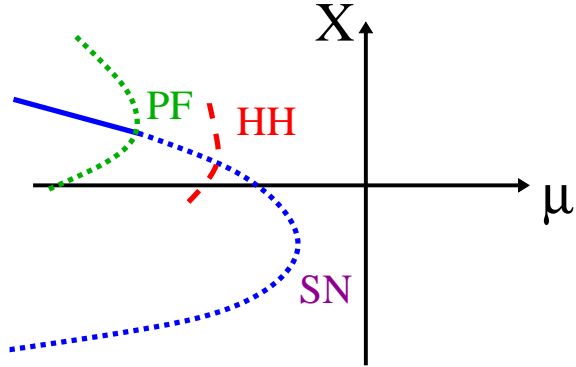,angle=0,width=3.0in}}
\caption{Sketch of the bifurcation diagram for a wave train
in a ring showing a stationary saddle-node bifurcation (SN) and a
subcritical pitchfork bifurcation (PF). In between these two
bifurcations is also a Hopf bifurcation (HH).}
\label{Fig-Pitchsketch}
\end{figure}

\noindent
This allows us to sketch the full bifurcation scenario for a wave
train in a periodic ring as depicted in
Figure~\ref{Fig-Pitchsketch}. In the subsequent sections we study the
bifurcations of the steady-state solution (\ref{statio}).

\medskip
\section{Bifurcation analysis of the Hopf bifurcation for a single pulse}
\label{Sec-BifHopf}

\noindent
In this Section we study the direction and the stability of the Hopf
bifurcation. In excitable media the Hopf bifurcation is a result of
stronger and stronger coupling of the activator with its own
inhibitor. Upon reducing the length of a ring for fixed excitability,
or reducing the excitability for a fixed ring length, a single pulse
will feel the tail of its own inhibitor created during its previous
passage through the ring. In a wave train each pulse will feel the
tail of the inhibitor of its neighbour in front. In the context of our
normal form (\ref{NF}) the increasing coupling translates into an
increase of $\beta \tau$. Upon increasing $\beta \tau$ from $\beta
\tau=1$ up to a critical value of $\beta \tau\approx 7.789$ we have
only one solution $\omega=\omega_0$ of the characteristic equation
(\ref{HomHopfa}); see Fig.~\ref{Fig-sine}. We will study now the
dynamics for this case. The case of arbitrary many solutions when one
encounters a pseudo-continuum of frequencies will be discussed further
down in Section~\ref{Sec-CGL}.\\

\noindent
The theory of bifurcation analysis for delay-differential equations is
well developed
\cite{Krasovskii,Hale,Hale2,Diekmann}. For example, in \cite{Diekmann}
a formula for the coefficients of the normal form for a Hopf
bifurcation is given explicitly. However, we found that the theory of
delay differential equations is not as well known amongst scientists
as its age may suggest. We find it therefore instructive to perform
the calculations explicitly and lead the reader through the
calculations.\\

\noindent
To study the direction of the Hopf bifurcation we have to
determine the sign of $d\sigma/d\mu$ at the bifurcation point
$\mu_H$. From (\ref{lin}) we infer
\begin{eqnarray*}
\frac{d\lambda}{d\mu}=-\frac{2g}{1-\beta \tau e^{-\lambda
\tau}}\frac{d{\bar X}}{d\mu}\; .
\end{eqnarray*}
Using (\ref{mueHopf}) we find from (\ref{statio})
\begin{eqnarray*}
\frac{d{\bar X}}{d\mu}_{|_{\mu_H}} =-\frac{1}{\beta(1-\cos{\omega \tau})}\; ,
\end{eqnarray*}
and subsequently
\begin{eqnarray}
\label{direction}
\frac{d\lambda}{d\mu}_{|_{\mu_H}} 
=\frac{2g}{\beta}\frac{1}{\|1-\beta \tau e^{-\lambda \tau}\|}>0\; ,
\end{eqnarray}
which implies that the stationary solution looses stability with
increasing values of the bifurcation parameter $\mu$.\\

\noindent
We now study the character of the Hopf bifurcation and derive the
normal form for a Hopf bifurcation from (\ref{NF}). In order to do
that we first transform the normal form (\ref{NF}) into standard form
by subtracting the stationary solution (\ref{statio}) according to
$X={\bar{X}}+x$ where ${\bar{X}}={\bar{X}}_1$. We obtain
\begin{eqnarray}
\label{nf}
\partial_t x = - 2g {\bar{X}} x - \beta (x(t-\tau) + \gamma_1x(t)) - gx^2\; ,
\end{eqnarray}
with the stationary solution being now $x(t)=0$. We will employ a
center manifold reduction for this equation to describe the dynamics
close to criticality. Center manifold theory is well-established for
maps, ordinary differential equations and partial differential
equations. However, although known for some time
\cite{Krasovskii,Hale,Hale2}, it is not well known how to formulate an
essentially infinite dimensional delay differential equation such as
(\ref{NF}) into a form such that center manifold reduction can be
applied. For ordinary differential equations, for example, the
application of center manifold theory is a straight-forward expansion
of the state vectors in critical eigenmodes. The problem for delay
differential equations is their inherent infinite dimensional
character. An initial condition $x(\theta)=x_0(\theta)$ for $-\tau \le
\theta \le 0$ is mapped onto a finite dimensional space; in the case
of (\ref{nf}) onto a $1$-dimensional space. Lacking uniqueness of
solutions is one obstacle which prohibits a straightforward
application of center manifold reduction. The trick out of this
dilemma is to reformulate the problem as a {\em mapping} from an
infinite-dimensional space of differentiable functions defined on the
interval $[-\tau,0]$, which we denote as
${\cal{C}}={\cal{C}}[[-\tau,0];\R]$ (i.e. $x_0(\theta)\in {\cal{C}}$),
onto itself. This allows us to employ the well established and
understood center manifold reduction for mappings. These ideas go back
to Hale \cite{Hale,Krasovskii}. We found well written examples of
center manifold reductions to be examined in
\cite{Wischert94,LeBlanc02,Krauskopf04}. In essence, the history of a
state vector $x(t)\in \R$ is folded to a single element of an extended
state space $x_t(\theta)\in {\cal{C}}$. In order to achieve this we
define $x_t(\theta)\in {\cal{C}}$ as
\[ x_t(\theta)=x(t+\theta)\quad {\rm for} \quad -\tau\le \theta \le 0
\; .\]
The time-evolution for $x(t)\in \R$ (\ref{nf}) needs to be
reexpressed in terms of propagators and operators acting on elements
of the extended state space $x_t(\theta)\in {\cal{C}}$ which can be
done by writing
\begin{eqnarray}
\label{XC}
\frac{d}{dt}x_t(\theta) = {\cal{A}}[x_t](\theta) = \left\{ 
\begin{array}{ll}
          \frac{d}{d\theta}x_t(\theta) & \mbox{if $-\tau\le \theta < 0$}\\
          {\cal{F}}[x_t] & \mbox{if $\theta=0$}
\end{array} 
\right.
\end{eqnarray}
For $-\tau\le \theta < 0$ we used the invariance condition
$dx(t+\theta)/dt=dx(t+\theta)/d\theta$. For $\theta=0$ we can split
the operator ${\cal{F}}$ into a linear part ${\cal{L}}$ and a
nonlinear part ${\cal{N}}$ and reformulate the right-hand side of
(\ref{nf}) as
\begin{eqnarray}
\label{F}
{\cal{F}}[x_t] = {\cal{L}}[x_t]+{\cal{N}}[x_t] \; ,
\end{eqnarray}
where 
\begin{eqnarray}
\label{FA}
{\cal{L}}[x_t] = \int_{-\tau}^0d\theta w_1(\theta) x_t(\theta) 
\quad {\rm with} \quad 
w_1(\theta) = - (2g{\bar{X}} + \beta\gamma_1)\delta(\theta)
- \beta \delta(\theta+\tau)
\end{eqnarray}
and
\begin{eqnarray}
\label{FN}
{\cal{N}}[x_t] = \int_{-\tau}^0d\theta_1d\theta_2
w_2(\theta_1,\theta_2) x_t(\theta_1)x_t(\theta_2) 
\quad {\rm with} \quad
w_2(\theta_1,\theta_2) = - g\delta(\theta_1)\delta(\theta_2)\; ,
\end{eqnarray}
where $\delta(\theta)$ denotes the Dirac $\delta$-function. Once
$x_t(\theta)$ is computed via solving (\ref{XC}), one may convert back
to $x(t)$ by \[ x(t)=\int_{-\tau}^0d\theta
\delta(\theta)x_t(\theta) \; .\]

\subsection{Linear eigenvalue problem}

We now linearize (\ref{XC}) around the stationary solution
$x_t(\theta)=0$ to obtain
\begin{eqnarray}
\label{XL}
\frac{d}{dt}\xi_t(\theta) = {\cal{A}}_L[\xi_t](\theta) 
= \left\{ 
\begin{array}{ll}
          \frac{d}{d\theta}\xi_t(\theta) & \mbox{if $-\tau\le \theta < 0$}\\
          {\cal{L}}[\xi_t] & \mbox{if $\theta=0$}
\end{array} 
\right.
\end{eqnarray}
%
%
%
The linear eigenvalue problem (\ref{XL}) can be solved using the
ansatz
\begin{eqnarray*}
\xi_t(\theta) = e^{\lambda t} \Phi(\theta) \quad {\rm for} \quad
-\tau\le \theta \le 0 \; .
\end{eqnarray*}
On the interval $-\tau\le \theta < 0$ we obtain 
\begin{eqnarray*}
\lambda \Phi(\theta) = \frac{d}{d\theta}\Phi(\theta)\; ,
\end{eqnarray*}
which is solved by
\begin{eqnarray}
\label{Phi}
\Phi(\theta) = e^{\lambda \theta}\Phi(0)\; .
\end{eqnarray}
Plugging the solution (\ref{Phi}) into (\ref{XL}) we can now evaluate
the $\theta=0$-part of (\ref{XL}) to obtain again the characteristic
equation (\ref{lin}). We recall the transcendental characteristic
equation as
\begin{eqnarray}
\label{LIN}
\lambda+2g{\bar X} + \beta \gamma_1 + \beta e^{-\lambda \tau} = 0\; .
\end{eqnarray}
Since in general the linear operator ${\cal{A}}_L$ is not selfadjoint,
we need to consider the corresponding adjoint eigenvalue problem on
the dual extended state space
${\cal{C}}^\dagger={\cal{C}}^\dagger[[0,\tau];\R] $. The dual problem
is given by backward evolution for $t\le0$,
i.e. $x_t^\dagger(s)=x_{-t}(-s)$ for $0\le s \le \tau$. The adjoint
problem can be formally written as
\begin{eqnarray*}
\frac{d}{dt}\xi^\dagger_t(s) = - {\cal{A}}^\dagger_L[\xi_t](s) \; .
\end{eqnarray*}
To provide an explicit form of the dual operator ${\cal{A}}_L$ we need
to define an inner product. It turns out that the normal scalar
product used for ordinary differential equations is not capable of
respecting the memory effects of delay-differential equations. The
following inner product is used
\begin{eqnarray}
\label{BiLin}
\langle\Psi^\dagger,\Phi \rangle =
\Psi^\dagger(0)\Phi(0)-\int_{-\tau}^0 d\theta \int_0^\theta ds
\Psi^\dagger(s-\theta) w_1(\theta) \Phi(s)
\; .
\end{eqnarray}
The adjoint operator is then explicitly given as
\begin{eqnarray}
\label{XLA}
{\cal{A}}^\dagger_L[\xi_t](s) = \left\{ 
\begin{array}{ll}
          -\frac{d}{ds}\xi^\dagger_t(s) & \mbox{if $0 < s \le \tau$}\\
          {\cal{L}}^\dagger[\xi^\dagger_t] & \mbox{if $s=0$}
\end{array} 
\right.
\end{eqnarray}
where 
\begin{eqnarray}
\label{LA}
{\cal{L}}^\dagger [\xi^\dagger_t] =  \int_{0}^\tau ds w_1(-s)
\xi^\dagger_t(s)
\; .
\end{eqnarray}
The adjoint eigenvalue problem (\ref{XLA}) is now solved using the
ansatz
\begin{eqnarray*}
\xi^\dagger_t(s) = e^{-\lambda t} \Psi^\dagger(s) \quad {\rm for} \quad
0< s\le \tau \; .
\end{eqnarray*}
On the interval $0 < s \le \tau$ we obtain 
\begin{eqnarray*}
-\lambda \Psi^\dagger(s) = \frac{d}{ds}\Psi^\dagger(s)\; ,
\end{eqnarray*}
which is solved by
\begin{eqnarray}
\label{Psi}
\Psi^\dagger(s) = e^{-\lambda s}\Psi^\dagger(0)\; .
\end{eqnarray}
Plugging the solution (\ref{Psi}) into (\ref{LA}) we can now evaluate
the $s=0$-part of (\ref{XLA}) to obtain again the characteristic
equation (\ref{LIN}). Note that the two solutions (\ref{Phi}) and
(\ref{Psi}) for the eigenvalue problem and its dual can be transformed
into each other by simple time-reversal $\theta\to -s$.

\noindent
Since the transcendental equation has two solutions with vanishing
real part $\sigma=0$, i.e. $\lambda=\pm i \omega$ with $\omega$ given
by (\ref{HomHopfa}), we have two solutions of the linear eigenvalue
problem (\ref{XL}) and its associated adjoint problem (\ref{XLA}). We
denote them as $\Phi_{1,2}$ and $\Psi^\dagger_{1,2}$ respectively. The
bilinear form (\ref{BiLin}) was constructed in order to assure
biorthogonality of the eigenfunctions. Defining the eigenfunctions as
\begin{eqnarray}
\label{EigFunc}
\Phi_1(\theta) = e^{i\omega \theta} \quad \Phi_2(\theta) = e^{-i\omega \theta}\\ 
\Psi_1^\dagger(s) 
= \nu e^{-i\omega s} \quad \Psi_2^\dagger(s) = \nu^\star e^{i\omega s} 
\end{eqnarray}
with the normalization constant
\begin{eqnarray}
\label{nu}
\nu = \frac{1}{1-\beta \tau e^{-i\omega \tau}}\; ,
\end{eqnarray}
we have $\langle\Psi^\dagger_i,\Phi_j \rangle = \delta_{ij}$ with
$i,j=1,2$ and $\delta_{ij}$ being the Kronecker symbol.\\

\subsection{Center-manifold theory}
\begin{figure}
\centerline{
\psfig{file=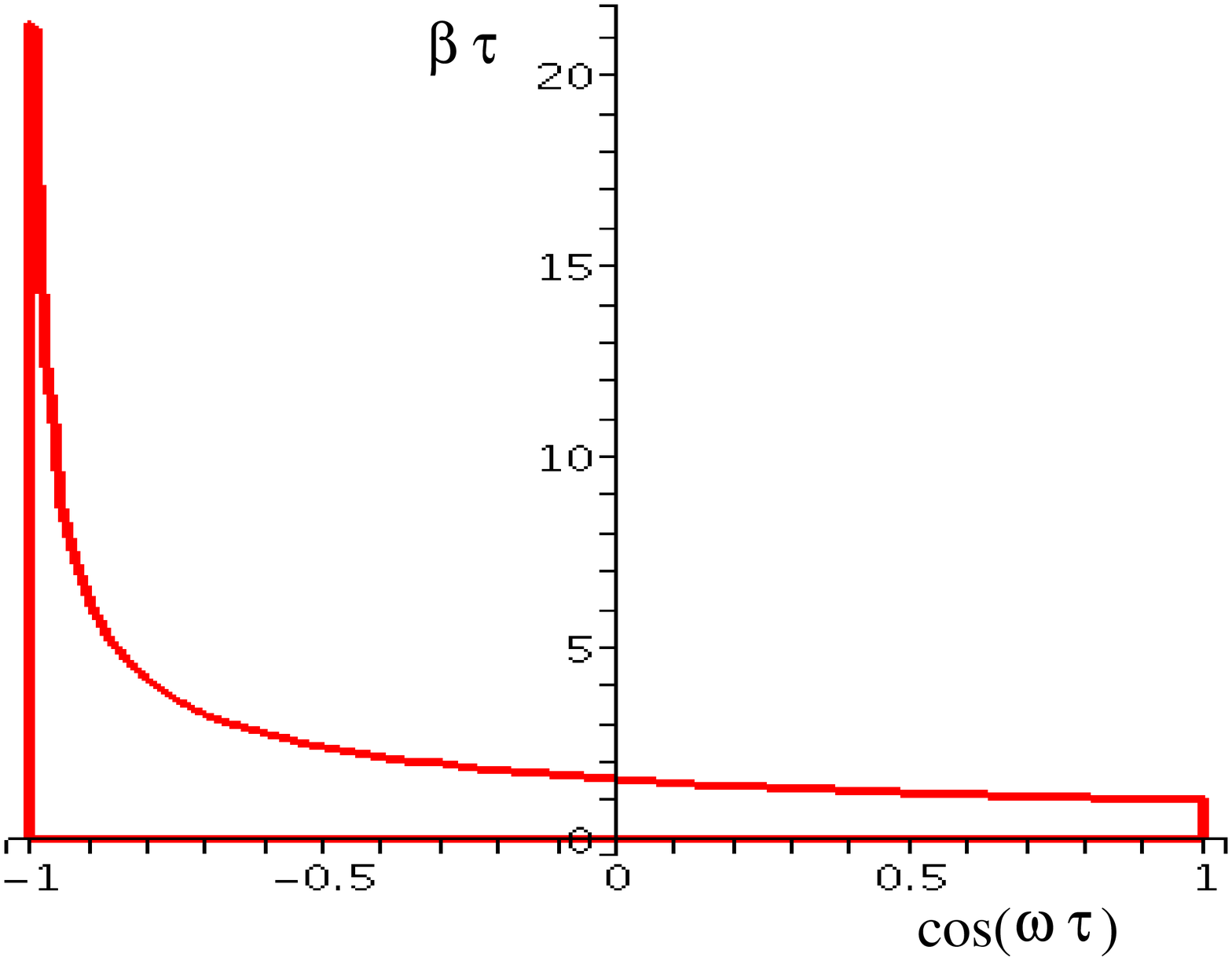,angle=0,width=3.0in}
\psfig{file=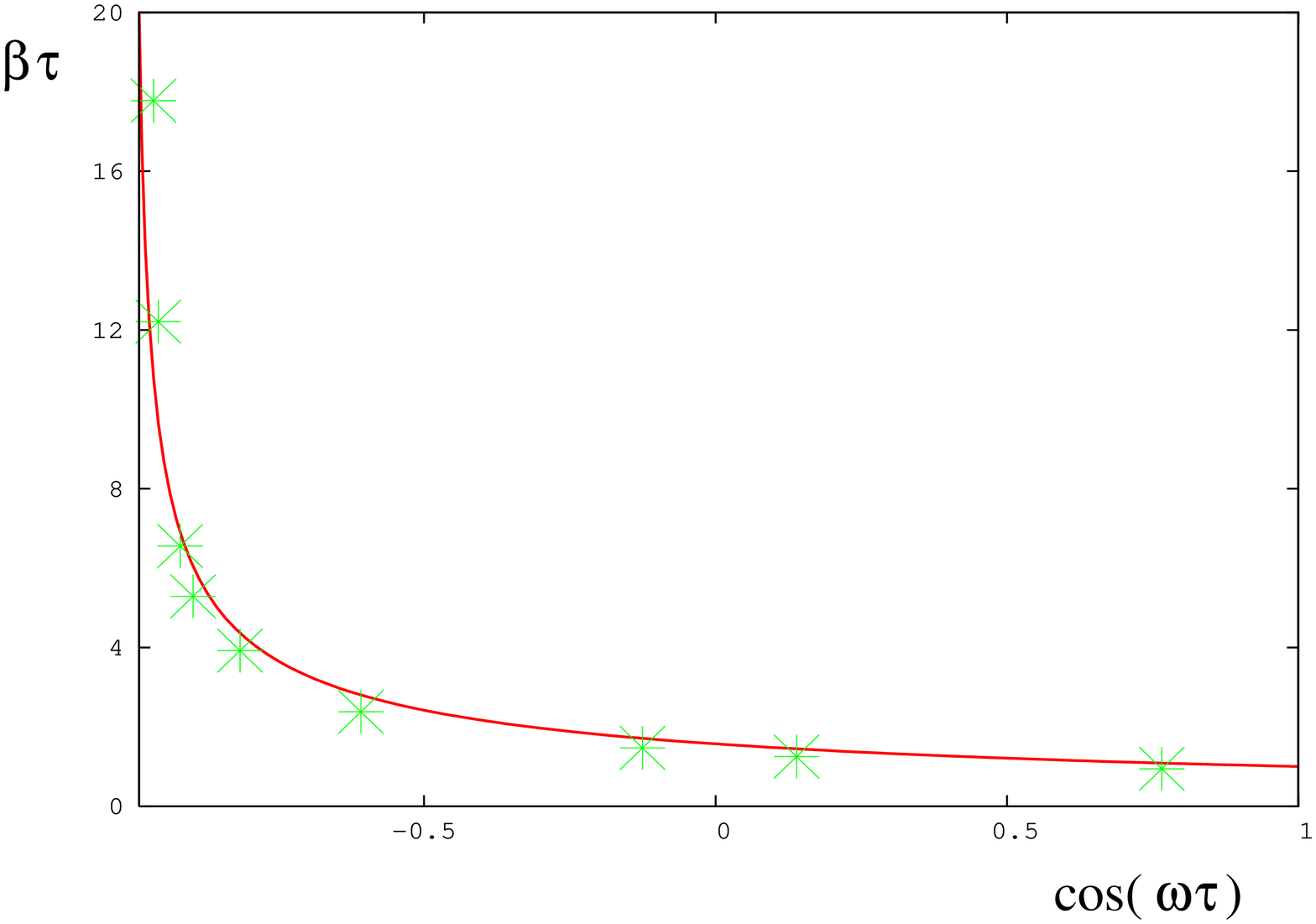,angle=0,width=3.0in}}
\caption{(a): Stability diagram for the equilibrium solution
(\ref{statio}). We also impose $\beta \tau\ge 0$. The region within
the bold lines is stable. Crossing the upper boundary corresponds to a
Hopf bifurcation. (b): Hopf line obtained by numerical simulations of
the modified Barkley model (\ref{barkley}) with $a=0.22$ and $u_s=0.1$
and $D=3$, \cite{Barkley91,GottwaldKramer04}. (Note that each point
corresponds to different values of $\epsilon$ and $L$). The numerical
results of the partial differential equations could be fitted to the
normal form (\ref{NF}) to obtain the parameters $\beta_0$ and $\tau$,
see Ref.~\cite{GottKram05}. The continuous line is the same Hopfline
as in (a).}
\label{Fig-Stabil}
\end{figure}

\noindent
For the nonlinear theory we need properties for the linear operator
${\cal{A}}$ developed in \cite{Hale,Hale2}. We summarize properties of
the transcendental characteristic equation (\ref{lin}): (i)
${\cal{A}}$ has a pure point spectrum, (ii) the real part of the
eigenvalues is bounded from above, and (iii) defining $a= \tau\{2g{\bar{X}}+\beta
\gamma_1\}$ and $b=\beta\tau$ all eigenvalues of
${\cal{A}}$ have negative real part if and only if (1.) $a>-1$, (2.) 
$a+b>0$ and (3.) $b<\zeta
\sin{\zeta}-a\cos{\zeta}$ where $\zeta$ is the root of
$\zeta=-a\tan{\zeta}$ with $0<\zeta<\pi$ for $a\neq 0$ and
$\zeta=\pi/2$ if $a=0$. These conditions can be translated for our
particular case (\ref{LIN}) using $a=-\beta\tau \cos(\omega
\tau)$. Condition (1) translates into $\beta \tau \cos(\omega\tau)<1$;
condition (2) into $\cos(\omega
\tau)<1$ and condition (3) defines a parameterized stability boundary
$\beta\tau<\zeta/\sin(\zeta)$ and $\cos(\omega
\tau)>\cos(\zeta)$ with $0<\zeta<\pi$. This last condition defines the
line in $\beta \tau$-$\cos(\omega\tau)$ space where the Hopf
bifurcation occurs. In particular we have $\beta\tau\ge 1$ and a
coalescence with the saddle node at $\cos(\omega \tau)=1$ with
$\beta\tau=1$ at the Bogdanov-Takens point. In Figure~\ref{Fig-Stabil}
we show the stability region. We note that care has to be taken in
interpreting the diagram in terms of the parameter $\tau$ because
$\beta=\beta(\tau)$ according to (\ref{beta}). In the limit $\tau\to
\infty$ we have $\beta \tau \to 0$. Note that stable solutions may
exist for $\beta \tau\ge 1$.

In \cite{Hale,Hale2} it is shown that under these circumstances one
may perform center manifold reduction. In particular we can decompose
$x_t(\theta)$ into slow modes associated with the eigenvalues
$\lambda=\pm i \omega$ and fast modes which correspond to modes with
negative real part of the eigenvalues. Center manifold theory says
that the fast modes are slaved to the slow modes and can be expressed
in terms of the slow modes. We therefore write
\begin{eqnarray}
\label{fastslow}
x_t(\theta)=z(t) \Phi_1(\theta) + {z^\star}(t) \Phi_2(\theta) +
h(z,z^\star)\; ,
\end{eqnarray}
where $z(t)\in\C$ and its complex conjugate $z^\star(t)$ are the
time-dependent amplitudes of the slow modes $\Phi_{1,2}(\theta)$ and
$h(z,z^\star)$ is the remaining fast component written as a function
of the slow amplitudes. The function $h(z,z^\star)$ is called the
center manifold. The expansion (\ref{fastslow}) is resemblant of
center-manifold theory for partial differential equations where
$\theta$ would be the spatial coordinate, and the expansion would be
an expansion of critical spatial eigenmodes. The connection between
delay differential equations and partial differential equations will
be explored further in Section~\ref{Sec-CGL}.\\

\noindent
We require the fast modes, and hence the center manifold, to lie in
the spectral complement of the centre space spanned by $\Phi(\theta)$;
we therefore have the constraint
\begin{eqnarray*}
\langle\Psi_j^\dagger,h(z,z^\star)\rangle=0\; \; \; \; \; \; j=1,2 \; .
\end{eqnarray*}
This implies for the slow amplitudes
\begin{eqnarray}
\label{slowamps}
z(t)=\langle\Psi_1^\dagger(\theta),x_t(\theta)\rangle 
\quad {\rm and} \quad 
z^\star(t)=\langle\Psi_2^\dagger(\theta),x_t(\theta)\rangle \; .
\end{eqnarray}
We use a near-identity transformation for the center manifold $h$ and
express it as a power series in $z$ and $z^\star$. The center manifold
is tangential to the manifold spanned by the slow modes which implies
the ansatz
\begin{eqnarray}
\label{CM}
h(z,z^\star)=\frac{1}{2}\left(h_{20}(\theta)z^2+2h_{11}(\theta) z
z^\star + h_{02}(\theta){z^\star}^2  \right) + {\cal{O}}(|z|^3) \; .
\end{eqnarray}
Since $x_t(\theta)$ is real we have
$h_{02}(\theta)=h^\star_{20}(\theta)$. Since the normal form for a
Hopf bifurcation only involves cubic terms, we only need to consider
quadratic terms in the equation for $h$ (\ref{ODEh}). The cubic terms
will then be generated via ${\cal{N}}[x_t](\theta=0)$ in the equations
for $z$ and $z^\star$ (see below, (\ref{ODEz}) and
(\ref{ODEzstar})).\\

\noindent
We will derive now ordinary differential equations for the slow
amplitudes $z$ and $z^\star$ which describe the dynamics on the slow
manifold. The theory of center manifolds tells us that the full
dynamics of (\ref{nf}) is well approximated by the slow dynamics
\cite{Carr}. The derivative of (\ref{slowamps}) with respect to time
$t$ is given by
\begin{eqnarray}
\label{ODEz}
{\dot{z}}&=&\langle \Psi_1^\dagger,{\dot{x}}_t\rangle \nonumber \\
&=& i \omega \langle \Psi_1^\dagger,x_t\rangle 
+ \langle \Psi_1^\dagger,{\cal{N}}[x_t]\rangle_{|_{\theta=0}} \nonumber \\
&=& i \omega z
+ \Psi_1^\dagger(0) {\cal{N}}[x_t](\theta=0) \nonumber \\
&=& i \omega z
+ \nu {\cal{N}}[x_t](\theta=0) \\
\label{ODEzstar}
{\dot{z}}^\star&=& -i \omega z^\star + \nu^\star {\cal{N}}[x_t](\theta=0)\\
{\dot{h}}&=&
{\dot{x_t}}-{\dot{z}}\Phi_1(\theta)-{\dot{z}}^\star\Phi_2(\theta)\nonumber\\
&=&{\cal{A}}_Lx_t+{\cal{N}}[x_t]-i\omega z\Phi_1(\theta)+i\omega
z^\star\Phi_2(\theta)-{\cal{N}}[x_t](\theta=0)\{\nu \Phi_1(\theta) +
\nu^\star\Phi_2(\theta)\} \nonumber\\
\label{ODEh}
&=&{\cal{A}}_Lh+{\cal{N}}[x_t]-\{\nu \Phi_1(\theta) +
\nu^\star\Phi_2(\theta)\}{\cal{N}}[x_t](\theta=0) \; ,
\end{eqnarray}
where the dot denotes a time derivative. Note that
${\cal{N}}[x_t]\neq0$ only for $\theta=0$ which can be written as
${\cal{N}}[x_t](\theta=0) = {\cal{N}}[z\Phi_1 + {z^\star} \Phi_2 +
h(z,z^\star)](\theta=0)$. Using (\ref{CM}) we have therefore
\begin{eqnarray}
\label{N}
{\cal{N}}[x_t](\theta=0) &=& -gx_t^2(\theta=0)\nonumber\\
&=&-g( z\Phi_1+z^\star\Phi_2+h)^2|_{\theta=0} \nonumber \\
&=& 
-g(z^2+{z^\star}^2
+ 2|z|^2 \nonumber \\
&& \hphantom{-g (}
+ h_{20}(0)z^3
+ (h_{20}(0)+2h_{11}(0)) |z|^2z
+ (h_{02}(0)+2h_{11}(0)) |z|^2z^\star
+ h_{02}(0){z^\star}^3)\nonumber\\
&& + {\cal{O}}(z^4,{z^\star}^4)\; .
\end{eqnarray}
Using the definition of ${\cal{A}}_L$ and ${\cal{N}}[x_t]$ we can
evaluate the evolution equation (\ref{XC}) by differentiating the
center manifold (\ref{CM}) with respect to time, equate with
(\ref{ODEh}), and obtain
\begin{eqnarray}
\label{ODE2}
{\dot{h}}= i\omega h_{20}(\theta)z^2 -i\omega
h_{02}(\theta){z^\star}^2 =
-2{\cal{R}}[\nu e^{i\omega \theta}]{\cal{N}}[x_t](\theta=0) +
\left\{
\begin{array}{ll}
          \frac{d}{d\theta}h(\theta) & \mbox{if $-\tau\le \theta < 0$}\\
          {\cal{H}}[z,z^\star;h] & \mbox{if $\theta=0$}
\end{array} 
\right.
\end{eqnarray}
where 
\begin{eqnarray}
{\cal{H}}[z,z^\star;h] = -(2g{\bar{X}}+\beta \gamma_1)h(0)-\beta
h(-\tau) + {\cal{N}}[x_t](\theta=0)\; .
\end{eqnarray}
Here ${\cal{R}}$ denotes the real part. 

\begin{figure}
\centerline{
\psfig{file=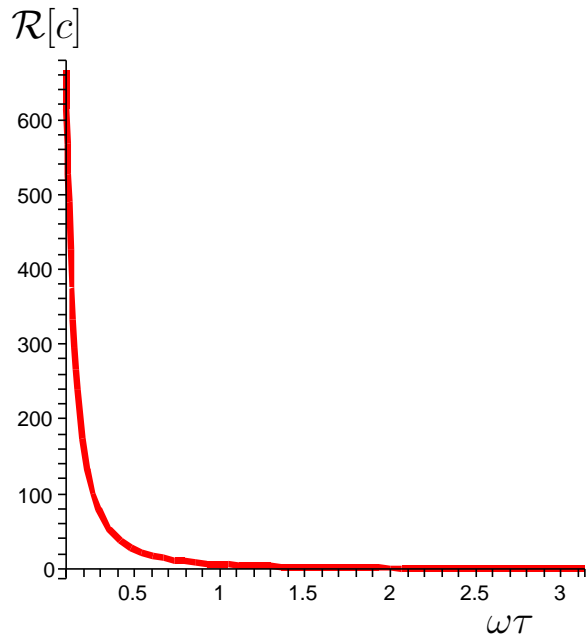,angle=0,width=3.0in}}
\caption{Plot of the real part of the cubic coefficient ${\cal{R}}[c]$ (\ref{cubicC})
as a function of $\omega \tau$. To produce the plot we set $g=1$ and
$\tau=1$. Both parameters are just coefficients multiplying $c$, so
they do not change the sign of $c$.}
\label{Fig-c}
\end{figure}

\noindent
Comparison of powers of $z$ and $z^\star$ yields differential
equations for $h_{ij}$ for the part with $-\tau\le\theta<0$ with an
associated boundary value problem coming from a comparison of powers
of $z$ and $z^\star$ from the $\theta=0$ part. We summarize
\begin{eqnarray}
\label{h20}
h_{20}^\prime&=&2i\omega h_{20}-4g{\cal{R}}[\nu e^{i\omega\theta}] \\
\label{h02}
h_{02}^\prime&=&-2i\omega h_{02}-4g{\cal{R}}[\nu e^{i\omega\theta}] \\
\label{h11}
h_{11}^\prime&=&-4g{\cal{R}}[\nu^{i\omega\theta}] \; ,
\end{eqnarray}
and the boundary conditions are given by
\begin{eqnarray}
\label{HBC1}
(i\omega -\beta e^{-i\omega\tau})h_{20}(0) + \beta h_{20}(-\tau) 
&=& 
2g(2{\cal{R}}[\nu]-1)\\
\label{HBC2}
(-i\omega -\beta e^{i\omega\tau})h_{02}(0) + \beta h_{02}(-\tau) 
&=& 
2g(2{\cal{R}}[\nu]-1)\\
\label{HBC3}
(-i\omega -\beta e^{-i\omega\tau})h_{11}(0) + \beta h_{11}(-\tau) 
&=& 
2g(2{\cal{R}}[\nu]-1) \; .
\end{eqnarray}
Note that the nonlinearity enters the differential equation in form of
an inhomogeneity. The ordinary differential equations
(\ref{h20})-(\ref{h11}) can be solved by variations of constants
\begin{eqnarray}
h_{20}(\theta)&=&H_{20}e^{2i\omega\theta}-2i\frac{g}{\omega}\left( \nu
e^{i\omega\theta}+\frac{1}{3}\nu^\star e^{-i\omega\theta} \right)\\
h_{02}(\theta)&=&H_{02}e^{-2i\omega\theta}+2i\frac{g}{\omega}\left( \nu^\star
e^{-i\omega\theta}+\frac{1}{3}\nu e^{i\omega\theta} \right)\\
h_{11}(\theta)&=&H_{11}+2i\frac{g}{\omega}\left( \nu
e^{i\omega\theta}-\nu^\star e^{-i\omega\theta} \right) \; .
\end{eqnarray}
The constants of integrations $H_{20}=H_{02}^\star$ and $H_{11}$ can
be determined using the boundary conditions
(\ref{HBC1})--(\ref{HBC3}). We obtain
\begin{eqnarray}
\label{HHH}
H_{20} &=& -\frac{2g}{i\omega -\beta e^{-i\omega\tau} +\beta
e^{-2i\omega\tau}} \\
H_{11} &=& -\frac{2g}{-i\omega -\beta e^{-i\omega\tau} +\beta} \; .
\end{eqnarray}
Note that $H_{11}=-2g/(\beta-\beta\cos(\omega \tau))=H_{11}^\star$. By
means of transformations \cite{Wiggins} equation (\ref{ODEz}) can be
transformed into the standard form for a normal form for Hopf
bifurcations
\begin{eqnarray}
{\dot{z}} = i \omega z + c |z|^2z\; .
\end{eqnarray}
The quadratic terms appearing in (\ref{ODEz}) can be eliminated by the
near-identity transformation $z \to z +
\eta_{20}z^2+\eta_{11}|z|^2+\eta_{02}z^{\star 2}$ using
$\eta_{20}=ig\nu/\omega$, $\eta_{11}=-2ig\nu/\omega$ and
$\eta_{02}=-ig\nu/3\omega$. Note that at the Bogdanov-Takens point
where $\omega=0$ such an elimination of quadratic terms is not
possible anymore. However this transformation generates further cubic
terms in (\ref{ODEz}). All of these cubic terms except those
proportional to $|z|^2z$ may be eliminated by means of another
transformation $z \to z + h_3(z,z^\star)$ where $h_3(z,z^\star)$ is a
cubic polynomial. After the transformation to eliminate the quadratic
terms the coefficient in front of the $|z|^2z$-term is found to be
\begin{eqnarray}
\label{cubicC}
c &=& -g \nu (h_{20}(0)+2h_{11}(0))\nonumber\\
&& 
+\left(2\eta_{02}g\nu^\star+2\eta_{11}g\nu^\star-\eta_{11}(i\omega
\eta_{20} + 3 g\nu)-2\eta_{20}(2i\omega \eta_{11}-g\nu) -2\eta_{02}g\nu \right) \nonumber \\
&=& -g \nu (h_{20}(0)+2h_{11}(0)) -
\frac{14}{3}ig^2\frac{|\nu|^2}{\omega}+\frac{20}{3}ig^2\frac{\nu^2}{\omega}
\nonumber \\ &=& -g \nu
(H_{20}+2H_{11}) + \frac{14}{3}ig^2\frac{\nu^2}{\omega} \; .
\end{eqnarray}
The stability and character of the Hopf bifurcation is determined by
the sign of the realpart of $c$. Because of (\ref{direction}) the Hopf
bifurcation is supercritical provided ${\cal{R}}[c]<0$ and subcritical
provided ${\cal{R}}[c]>0$. These criteria can be easily deduced by
writing $z=re^{i\phi}$. Note that ${\cal{R}}[c]$ can be written as a
function of $g,\tau$ and $\omega$ only since $\beta=\omega/\sin(\omega
\tau)$ at the Hopf bifurcation. Using algebraic software packages such
as Maple, we can show that ${\cal{R}}[c]>0$ for all values of $g,\tau$
and $\beta$. In Figure~\ref{Fig-c} we show the real part of $c$ as a
function of $\omega \tau$. This confirms that the Hopf bifurcation is
indeed as conjectured in \cite{GottKram05} subcritical. We have
checked our result against numerical simulations of the full
normal-form (\ref{NF}) and also using the software package DDE-BIFTOOL
\cite{ddebiftool}. Again, the degeneracy at $\omega \tau \to 0$ is
reflected in ${\cal{R}}[c]$ by a singularity at $\omega \tau=0$.

We will discuss the implications of this result in
Section~\ref{Sec-Disc}.


\medskip
\section{The limit of large delay times: The Hopf bifurcation for wave
trains}
\label{Sec-CGL}
\begin{figure}
\centerline{
\psfig{file=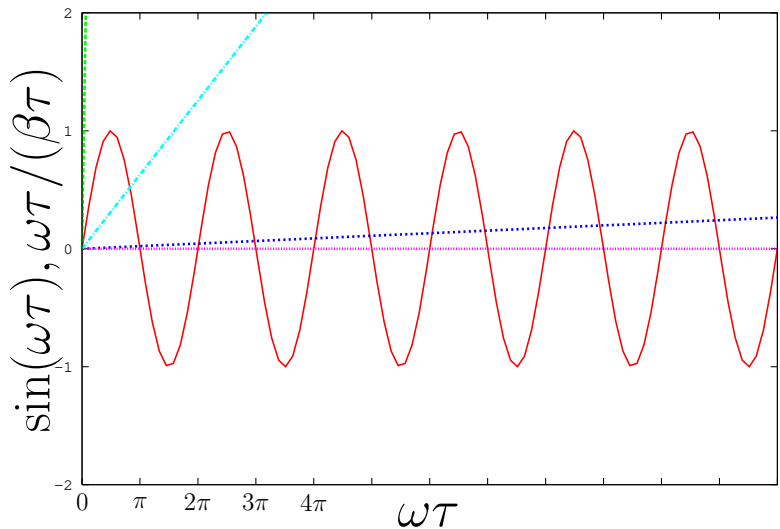,angle=0,width=4.0in}}
\caption{Illustration of the solutions and number of solutions of the
implicit equation (\ref{omegasin}) for $\omega$. Green curve:
$\beta\tau=0.1$, no Hopf bifurcation; light blue curve: $\beta\tau=5$,
Hopf bifurcation with one marginal mode; dark blue curve:
$\beta\tau=143$, Hopf bifurcation with finitely many marginal modes;
pink curve: $\beta\tau=\infty$, Hopf bifurcation with infinitely many
marginal modes.}
\label{Fig-sine}
\end{figure}

In the previous Section we have described the Hopf bifurcation for
small $\omega \tau$ when there is only one marginal mode. This
describes the behaviour of a single pulse on a ring. In this Section
we will pursue the case of large delay times when a pseudo-continuum
of critical Hopf modes occurs. We will derive a Ginzburg-Landau
equation as an amplitude equation describing near-threshold behaviour
of such a pseudo-continuum. The connection between amplitude equations
and delay-differential equations has long been known
\cite{Giacomelli96,Giacomelli98,Schanz03,Nizette03,Amann06,WolframYanchuk}. The
cross-over from a finite-dimensional center-manifold to an
infinite-dimensional amplitude equation can be best viewed when
looking at the Hopf condition (\ref{HomHopfa})
\begin{eqnarray}
\label{omegasin}
\omega=\beta\sin{\omega \tau}\; .
\end{eqnarray}
For $\beta \tau > 7.789$ there are at least two solutions of
(\ref{omegasin}) for $\omega$. This equation has arbitrary many
solutions $\omega_k$ for $\beta \tau
\to \infty$ and we obtain a pseudo-continuum in an interval with lower
closed boundary at $\omega \tau$ and upper boundary $\omega \tau =
\pi$. At the singular limit $\omega \tau=\pi$ there are countably
infinitely many eigenvalues $\omega_k \tau = k
\pi$. An illustration is given in Fig.~\ref{Fig-sine}. Note that the
upper boundary $\omega \tau = \pi$ corresponds to the coalescence of
the Hopf bifurcation with the pitchfork bifurcation in the case of
several pulses on a ring (see Section~\ref{Sec-NF}).

All these solutions are marginal and would have to be included in the
ansatz (\ref{fastslow}). Note that for excitable media where
$\beta=\beta_0 \exp(-\kappa \tau)$ (see (\ref{beta})) this limit
cannot be achieved by simply letting $\tau\to \infty$. The function
$\beta \tau$ has a maximum at $\tau=1/\kappa$. So in order to have
$\beta \tau \to \infty$ one can either have $\beta_0\to \infty$ which
seems unphysical or $\kappa\to 0$ with $\tau\to \infty$ to keep
$\kappa \tau$ finite. Hence the limit $\beta \tau\to \infty$ applies
to media with a very slowly decaying inhibitor in a very large
domain. Large domain instabilities are known from certain excitable
reaction diffusion systems in the context of autocatalytic oxidation
of $CO$ to $CO_2$ on platinum \cite{Baer1,Baer2,Baer3}.\\

\noindent
The case $\omega \tau \approx \pi$ is important for single pulses in a
ring and for wave trains. Firstly, it describes the case for a single
pulse when a continuum of modes becomes unstable to a Hopf
bifurcation. But more importantly it describes the case of a wave
train with distinct members when the pitchfork bifurcation coalesces
with the Hopf bifurcation. The point of coalescence is at $\mu_{PF}$
given by (\ref{AI_PFa}) and with amplitude given by (\ref{AI_PFb})
which we recall
\[
X_{PF}=\frac{\beta}{2g}(1-\gamma_1)\; .
\] 
At this point of coalescence the Hopf frequency is in resonance with
the spatial instability in which every second pulse dies. At
$\mu_{PF}$ the two equations (\ref{NF2}) describing the alternating
modes in a wave train collapse to the single equation for one pulse
(\ref{NF}) (see also Fig.~\ref{Fig-Pitchsketch}). This Section will
investigate whether the coalescence of the subcritical pitchfork
bifurcation with the Hopf bifurcation may produce stable
oscillations.\\

\noindent
We will perform a multiple scale analysis of the normal form
(\ref{NF}) along the lines of \cite{Nizette03}. We will obtain at
third order an evolution equation for the amplitude as a solvability
condition which describes the dynamics close to the Hopf
bifurcation. We consider the case of large delay times $\tau$ and
introduce a small parameter $\epsilon=1/\tau$. To capture the dynamics
close to the point of coalescence we introduce a slow time scale
\[
s=\epsilon t \; ,
\]
and rewrite the normal form (\ref{NF}) in terms of the slow variable
as
\begin{eqnarray}
\label{NFGL}
\epsilon \partial_s X = -\mu - g X^2 - \beta (\gamma + X(s-1) + \gamma_1X)\; .
\end{eqnarray}
We expand the scalar field $X(s)$ as
\[
X=x_{PF} + \epsilon x_1 + \epsilon^2 x_2 + \epsilon^3 x_3 + \cdots\; .
\]
Using the generic scaling the bifurcation parameter can be written as
\[
\mu=\mu_{PF}+\epsilon^2\Delta \mu  + \cdots\; .
\]
A Taylor expansion of (\ref{omegasin}) around $\omega \tau=\pi$ yields
at first order $\omega \tau=\beta \tau(\pi-\omega \tau)$ which for
large $\tau$ (small $\epsilon$) we may write as
\begin{equation}
\label{e-omegataupi}
\omega \tau = \pi(1-\frac{1}{\beta \tau}+\frac{1}{(\beta \tau)^2}) =
\pi(1-\frac{1}{\beta}\epsilon + \frac{1}{\beta^2}\epsilon^2)\; .
\end{equation}
This suggests a multiple time scaling
\[
\partial_s=\partial_{s_0}+\epsilon \partial_{s_1}+\epsilon^2
\partial_{s_2}  + \cdots\; .
\]
Close to the bifurcation point critical slowing down occurs which
allows us to expand the delay term for large delays as
\begin{eqnarray}
X(s-1)&=&e^{-\partial_{s}}X(s)\\
&\approx&\left[1 - \epsilon\partial_{s_1} +
\epsilon^2\left(\frac{1}{2}\partial_{s_1s_1} -
\partial_{s_2} \right)
\right] e^{-\partial_{s_0}}X(s) \; .
\end{eqnarray}

\noindent
At lowest order, ${\cal{O}}(1)$, we obtain the equation determining
$x_{PF}$. At the next order we obtain
\begin{eqnarray}
\label{e-firstO}
{\cal{L}}x_1=0 \;,
\end{eqnarray}
with the linear operator
\begin{eqnarray*}
{\cal{L}} = \beta \left[1+e^{-\partial_{s_0}}\right]\; .
\end{eqnarray*}
Equation (\ref{e-firstO}) is solved by
\begin{eqnarray}
\label{e-x1}
x_1(s_0,s_1,s_2)=z(s_1,s_2) e^{i\pi s_0} + {\bar{z}}(s_1,s_2) e^{-i\pi
s_0}\; ,
\end{eqnarray}
with complex amplitude $z$ and its complex conjugate ${\bar{z}}$. Note
that on the fast time scale $t$ we would have $x_1(t)=z\exp(i\omega
t)+ {\rm c.c.}$ with $\omega \tau = \pi$, which, of course, is the
Hopf mode at onset.\\

\noindent
At the next order,
${\cal{O}}(\epsilon^2)$, we obtain
\begin{eqnarray}
\label{e-secondO}
{\cal{L}}x_2=-\Delta\mu-gx_1^2-\partial_{s_0}x_1 + \beta
\partial_{s_1}e^{-\partial_{s_0}}x_1\; .
\end{eqnarray}
The right-hand side involves terms proportional to $\exp(\pm i \pi
s_0)$, which are resonant with the homogeneous solution of
${\cal{L}}x_2=0$. We therefore impose the solvability condition
\begin{eqnarray*}
\partial_{s_0}x_1 -\beta\partial_{s_1}e^{-\partial_{s_0}}x_1 = 0\; ,
\end{eqnarray*}
which using (\ref{e-firstO}) reads as
\begin{eqnarray}
\label{e-cg0}
\partial_{s_1}x_1 + \frac{1}{\beta}\partial_{s_0}x_1 = 0\; .
\end{eqnarray}
In terms of the complex amplitude $z$ using (\ref{e-x1}) this reads as
\begin{eqnarray}
\label{e-cg}
\partial_{s_1}z + \frac{1}{\beta}i\pi z = 0\; .
\end{eqnarray}
This amounts to the time scale $i\omega \tau \approx \tau\partial_t
\approx \partial_{s_0}+\epsilon \partial_{s_1} = i\pi -
i\epsilon\pi/\beta=i\pi(1-1/(\beta \tau))$ which corresponds to our
scaling (\ref{e-omegataupi}) at first order. Provided (\ref{e-cg}) is
satisfied we can readily solve (\ref{e-secondO}) by solving for each
appearing harmonic, and find
\begin{eqnarray}
\label{e-x2}
x_2&=&-\frac{1}{2\beta}\left[ \Delta\mu + 2g|z|^2 
+ gz^2e^{2i\pi s_0}
+ g{\bar{z}}^2e^{-2i\pi s_0} \right]
\nonumber
\\
&=&-\frac{1}{2\beta}\left[ \Delta\mu + gx_1^2 \right]
\; ,
\end{eqnarray}
where we used (\ref{e-firstO}).

\noindent
At the next order, ${\cal{O}}(\epsilon^3)$, we obtain the desired
evolution equation as a solvability condition. At
${\cal{O}}(\epsilon^3)$ we obtain
\begin{eqnarray}
\label{e-third0}
{\cal{L}}x_3&=&-\partial_{s_0}x_2 + \beta
\partial_{s_1}e^{-\partial_{s_0}} x_2  
- \partial_{s_1}x_1 - 2gx_1x_2 - \frac{1}{2}\beta
\partial_{s_1s_1} e^{-\partial_{s_0}} x_1 + \beta \partial_{s_2}
e^{-\partial_{s_0}} x_1
\nonumber \\
&=&
-\partial_{s_0}x_2 + \beta
\partial_{s_1}e^{-\partial_{s_0}} x_2 - \partial_{s_1}x_1 
- 2gx_1x_2  + \frac{1}{2}\beta \partial_{s_1s_1} x_1 
- \beta \partial_{s_2}
x_1
\; .
\end{eqnarray}
Again resonant terms proportional to $\exp(\pm i \pi s_0)$ are
eliminated by imposing a solvability condition which upon using the
expressions for $x_2$ yields the desired amplitude equation
\begin{eqnarray}
\label{GL}
\partial_{s_2} x_1 -\frac{1}{\beta^2}\partial_{s_0} x_1
=  
\frac{g}{\beta^2} \Delta \mu \,x_1
+ \frac{1}{2\beta^2}\partial_{s_0 s_0}x_1 
+ \frac{g^2}{\beta^2}x_1^3\; .
\end{eqnarray}
This is the well-studied real Ginzburg-Landau equation
\cite{KramerGL}. The time-like variable is the slow time scale $s_2$
and the space-like variable the faster time scale $s_0$ which is
${\cal{O}}(\tau)$. As in the finite dimensional case studied in
Section~\ref{Sec-BifHopf} the Hopf bifurcation is clearly subcritical
since the real part of the coefficient in front of the cubic term in
(\ref{GL}) is positive for all parameter values. Hence the coalescence
of the Hopf bifurcation and the pitchfork bifurcation cannot lead to
stable oscillations. We have shown that wave trains also undergo
unstable oscillations in the framework of the normal form
(\ref{NF}).\\

The usefulness of the spatio-temporal view point for delay
differential equations as expressed here in the Ginzburg-Landau
equation (\ref{GL}) has been pointed out
\cite{Giacomelli96,Giacomelli98,Nizette03,WolframYanchuk}. However the
Ginzburg-Landau equation (\ref{GL}) may be cast into a finite
dimensional system which emphasizes the underlying multiple scale
analysis. We start by rewriting (\ref{GL}) as an equation for the
complex amplitude $z$. One can explicitly express $s_0$-derivatives
and obtain the following finite dimensional system
\begin{eqnarray}
\label{GLz}
\partial_{s_2} z -i\pi \frac{1}{\beta^2} z
=  
\frac{g}{\beta^2} \left(
\Delta \mu 
-\frac{\pi^2}{2 g}
\right)\,z
+ 3\frac{g^2}{\beta^2}|z|^2z\; .
\end{eqnarray}
The time-scaling on the left hand-side is as expected from our initial
linearization and expansion of the frequency (\ref{e-omegataupi}). We
have in total
\[
i\omega \tau\approx\tau \partial_t=\partial_s\approx
\partial_{s_0}+\epsilon\partial_{s_1}+\epsilon^2\partial_{s_2}=i\pi\left(
1-\frac{1}{\beta \tau}+\frac{1}{(\beta\tau)^2} \right)\;,
\]
which corresponds to (\ref{e-omegataupi}). This illustrates the
multiple-scale character of our analysis where the nonlinear term may
be interpreted as a frequency correction \cite{F00}. The correction
term to the linear term on the right-hand side of (\ref{GLz}) shows
that the onset is retarded on the very slow time scale $s_2$.\\

\noindent
In \cite{Echebarria02} a real Ginzburg-Landau equation was derived for
paced excitable media with an additional integral term modeling the
pacing. It would be interesting to see whether the therein derived
amplitude equation can be derived in a multiple scale analysis along
the lines of this multiple scale analysis.\\


\medskip
\section{Summary and Discussion}
\label{Sec-Disc}

\noindent
We have explored the Hopf bifurcations of a single pulse and of a wave
train in a ring of excitable medium. We have found that for the
phenomenological normal form (\ref{NF}) the Hopf bifurcation for a
single pulse on a ring and for a wave train on a ring is always
subcritical independent on the equation parameters.\\

\noindent
Hopf bifurcations in excitable media had been previously
studied. Besides numerical investigations of the Barkley model
\cite{Knees92}, the modified Barkley model \cite{GottwaldKramer04},
the Beeler-Reuter model
\cite{BeelerReuter,QuanRudy,Courtemanche,Karma94,Courtemanche96,Vinet00}, the
Noble-model \cite{Noble,Karma93,Karma94} and the Karma-model \cite{Karma93},
where a Hopf bifurcation has been reported, there have been many
theoretical attempts to quantify this bifurcation for a single-pulse
on a ring. Interest has risen recently in the Hopf bifurcation in the
context of cardiac dynamics because it is believed to be a precursor
of propagation failure of pulses on a ring. The Hopf bifurcation has
been related to a phenomenon in cardiac excitable media which goes
under the name of {\it alternans}. Alternans describe the scenario
whereby action potential durations are alternating periodically
between short and long periods. The interest in alternans has risen as
they are believed to trigger spiral wave breakup in cardiac tissue and
ventricular fibrillation
\cite{Nolasco,Courtemanche,Karma93,Karma_A,Fenton02}.\\

\noindent
Our results may shed a new light on what may be called {\it
alternans}. The occurrence of alternans in clinical situations is
often followed by spiral wave breakup and ventricular fibrillation
\cite{Nolasco,Courtemanche,Karma_A,Karma93,Fenton02}. The subcritical
character of the Hopf bifurcation gives a simple and straightforward
explanation for this phenomenon. Moreover, if the system length $L$ is
slowly varied, long transients may be observed of apparently stable
oscillations (see Figure~\ref{Fig-karma} and
Figure~\ref{Fig-karma2}). Depending on whether the system length is
below or above the critical length $L_H$ the oscillations will relax
towards the homogeneous state or the instability will lead to wave
breakup. However, even for the case of relaxation towards the stable
homogeneous solution, these oscillations may lead to wave breakup upon
further reduction of the system length, because of the subcritical
character of the Hopf bifurcations. This illustrates the diagnostic
importance of cardiac alternans.\\


\subsection{Limitations and range of validity of our results}

\noindent
Strictly speaking, our result that the Hopf bifurcation is subcritical
for the normal form (\ref{NF}) cannot be taken as a prove that
alternans are unstable for all excitable media. The normal form
(\ref{NF}) is only valid for a certain class of excitable media. In
particular it describes the situation in which an activator weakly
interacts with the inhibitor of the preceding exponentially decaying
inhibitor. Moreover, the normal form has only been phenomenologically
derived in \cite{GottKram05}. Of course, unless a rigorous derivation
of the normal form (\ref{NF}) has been provided the results presented
here may serve as nothing more than a guidance in interpreting
alternans in real cardiac systems or more complex ionic models of
excitable media, and may alert scientists to check results on
stability of oscillations more carefully.\\

\noindent
Several simplifications have been made to obtain the normal form
(\ref{NF}) in \cite{GottKram05}. For example, the time delay
$\tau=L/c_0$ is treated as constant. This is obviously not correct for
Hopf bifurcations.
However, the inclusion of $\gamma_1$ (which is essential in the
quantitative description of the Hopf bifurcation) allows for velocity
dependent effects. Guided by the success of the normal form to
quantitatively describe a certain class of excitable media and by
numerical experiments we are hopeful that our result may help
interpreting experiments and numerical simulations.\\

\noindent
In Section~\ref{Sec-numerics} we will discuss a particular model for
cardiac dynamics in which for certain parameter values the assumptions
for the derivation of our normal form are violated. For these
parameter values stable oscillations may occur. However even for
systems which are described by the normal form (\ref{NF}) a word of
caution is appropriate. If the oscillatory solutions bifurcating from
the stationary solution are unstable as we have proven here, the
unstable Hopf branch could in principle fold back and restabilize. Our
analysis does not include such secondary bifurcations. Another
scenario which we cannot exclude based on our analysis is that the
unstable branch may be a basin of attraction for a stable oscillatory
solution far away from the homogeneous solution. However, our
numerical simulations do not hint towards such scenarios.\\ From an
observational perspective the relevance of the subcritical instability
for spiral wave breakup is a matter of the time scale of the
instability. The time scale associated with the subcritical Hopf
bifurcation may be very long as seen in Fig.~\ref{Fig-barkley}. This
time scale becomes shorter the further the perturbation in the
bifurcation parameter is from its value at the corresponding stable
stationary pulse solution. In any case, if the parameter is kept fixed
above the critical value, the instability will eventually develop
unless the life time of a reentrant spiral is less than the time scale
of the instability. For clinical applications one would need to
estimate the time scale of a reentrant spiral and compare it with the
time scale of the instability. Such estimates however are not
meaningful for simple models such as the Barkley model.\\

\noindent 
Our definition of alternans is restricted to non-paced pulses on a
ring. If the excitable media is paced, the subcritical character of
the Hopf bifurcation is not guaranteed anymore, and there is no {\it a
priori} reason why stable alternans cannot occur. Indeed, in
periodically stimulated excitable media stable alternans have been
reported
\cite{Guevara81,Lewis90,Hastings00,Guevara02,Echebarria02,Fox02,Henry05}.
A non-paced single pulse on a ring is a simple model for a reentrant
spiral moving around an anatomical obstacle or around a region of
partially or totally inexcitable tissue. As such it ignores the
dynamics of the spiral away from the obstacle. An extension would be
to look at a transversal one-dimensional slice through a spiral and
consider wave trains and instabilities of such wave trains.


\subsection{Relation to the restitution condition}

Since the pioneering work \cite{Guevara84} alternans have been related
to a period-doubling bifurcation. This work has rediscovered the
results by \cite{Nolasco}, which had hardly been noticed by the
scientific community until then. In there it was proposed that the
bifurcation can be described by a one-dimensional return map relating
the action potential duration ($APD$) to the previous recovery time,
or diastolic interval ($DI$), which is the time between the end of a
pulse to the next excitation. A period-doubling bifurcation was found
if the slope of the so called restitution curve which relates the
$APD$ to the $DI$, exceeds one. A critical account on the predictive
nature of the restitution curve for period-doubling bifurcations is
given in
\cite{Fenton99,Fox02}. In \cite{Karma94} the instability was analyzed
by reducing the partial differential equation describing the excitable
media to a discrete map via a reduction to a free-boundary problem. In
\cite{GottwaldKramer04} the Hopf bifurcation could be described by
means of a reduced set of ordinary-differential equations using a
collective coordinate approach. In
\cite{Courtemanche,Courtemanche96,Vinet00,Vinet03} the bifurcation was
linked to an instability of a single integro-delay equation. The
condition for instability given by this approach states - as in some
previous studies involving one-dimensional return maps - that the
slope of the restitution curve needs to be greater than one. However,
as evidenced in experiments \cite{Hall,Banville02} and in theoretical
studies \cite{Fenton99,Fox02,Keener02,Cherry04,Bauer07} alternans do
not necessarily occur when the slope of the restitution curve is
greater than one. In our work we have a different criterion for
alternans (which we interpret now as unstable periodic
oscillations). Our condition for the occurrence of alternans, $\beta
\tau > 1$, does not involve the restitution curve but involves the
coupling strength and the wave length. Moreover, in
Fig.~\ref{Fig-Stabil}(b) we can see that for our normal form pulses
can be stable for values of $\beta \tau \gg 1$ in accordance with the
above mentioned experiments and numerical studies.\\ 

\noindent
In the following we will show how our necessary condition for the
onset of instability $\beta \tau > 1$ can be related to the
restitution condition, that the onset of instability is given when the
slope of the restitution curve exceeds $1$.\\

\noindent
Close to the saddle node the Hopf frequency is $\omega \tau
\approx 0$. We introduce a small parameter $\delta\ll1$ and write
close at the saddle node 
\[
X={\bar X}_{SN} + \delta x \; ,
\]
where ${\bar X}_{SN}$ is given by (\ref{SN_wt}). The generic scaling
close to the saddle node implies that we may write
$\mu=\mu_{SN}+\delta^2 \Delta \mu$. Using the critical slowing down at
the saddle node and the fact that $\omega \tau \approx 0$ we may
approximate the normal form (\ref{NF}) to describe the temporal change
of $X$ at some time $t$ and at some later time $t+\tau$.
\begin{eqnarray*}
\label{e-backwardeuler}
\frac{\delta x_{n+1}-\delta x_n}{\tau} = -\mu_{SN}-\delta^2\Delta \mu
- g({\bar X}_{SN} + \delta x_n)^2 - \beta(\gamma + {\bar X}_{SN} +
\gamma_1 {\bar X}_{SN} + \delta x_{n-1} + \gamma_1 \delta x_{n})\; .
\end{eqnarray*}
Here $x_n=x(t_n)$ and $x_{n+1}=x(t_n+\tau)$. Neglecting terms of
${\cal{O}}(\delta^2)$ and using the definition of the saddle node
(\ref{SN_wt}) we end up with
\begin{eqnarray*}
\label{e-backwardeuler2}
x_{n+1}-(1+\beta \tau) x_n + \beta \tau x_{n-1} = 0\; .
\end{eqnarray*}
This equation has either the solution $x_n=1$ which corresponds to the
stable steady solution described by ${\bar X}_1$ of (\ref{statio}), or
\begin{eqnarray*}
x_{n}= (\beta \tau)^n x_0\;,
\end{eqnarray*}
which implies
\begin{eqnarray}
\label{e-restcond}
x_{n}= \beta \tau x_{n-1}\; .
\end{eqnarray}
Close to the saddle node the amplitude of the activator correlates
well with the APD, and we find that $\beta \tau > 1$ is exactly the
restitution condition whereby the slope of the restitution curve has
to be larger than one.\\

\noindent
Our model contains the restitution condition as a limiting case when
the Hopf bifurcation occurs close to the saddle node. However, as seen
in Fig.~\ref{Fig-Stabil} $\beta \tau$ may be larger than one but still
the system supports stable pulses. These corrections to the
restitution conditions are captured by our model. Moreover, the normal
form is able to determine the frequency at onset.\\

\noindent
We note that the parameter $\gamma_1$ does not enter the restitution
condition; it is not needed for the {\it existence} of a Hopf
bifurcation (cf. (\ref{HomHopfa}) and (\ref{HomHopfb})). However, as
pointed out in \cite{GottKram05} quantitative agreement with numerical
simulations is only given if $\gamma_1$ is included. In
\cite{GottKram05} the inclusion of the $\gamma_1$-term takes into
account the velocity dependent modifications of the bifurcation
behaviour: large-amplitude pulses have a higher velocity than
low-amplitude ones. A larger pulse will therefore run further into the
inhibitor generated by its predecessor. Velocity restitution curves
have been studied in
\cite{Keener02} to allow for a modification of the restitution
condition derived in \cite{Courtemanche96} for a single pulse in a
ring. The normal form incorporates naturally these velocity dependent
terms.\\

\noindent
For a recent numerical study on the validity of the restitution
condition the reader is referred to \cite{Bauer07}. In this work the
stability of certain excitable media is investigated by means of
numerical continuation methods which allows a precise identification
of the onset of oscillations. At the onset of alternans the
restitution curve was determined. It was found that the restitution
condition failed for three out of four cases for pulses in a
one-dimensional ring. Our result suggests that the restitution
condition may be a good indicator for the onset of alternans close to
the saddle node.\\


\subsection{Numerical simulations}
\label{Sec-numerics}
\begin{figure}
\centerline{
\psfig{file=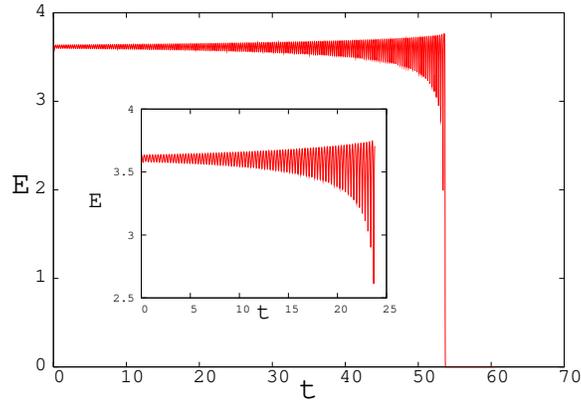,angle=270,width=3.0in}}
\caption{Temporal behaviour of the maximal amplitude $E_{max}$ of
the activator $E$ for model (\ref{karma}) just above the subcritical
Hopf bifurcation. The parameters are $A=1.5415$, $\epsilon=0.009$,
$M=30$ and $n_B=0.525$ and $L=0.215$. The inlet shows the behaviour at
$L=0.210$.}
\label{Fig-karma}
\end{figure}

\begin{figure}
\centerline{
\psfig{file=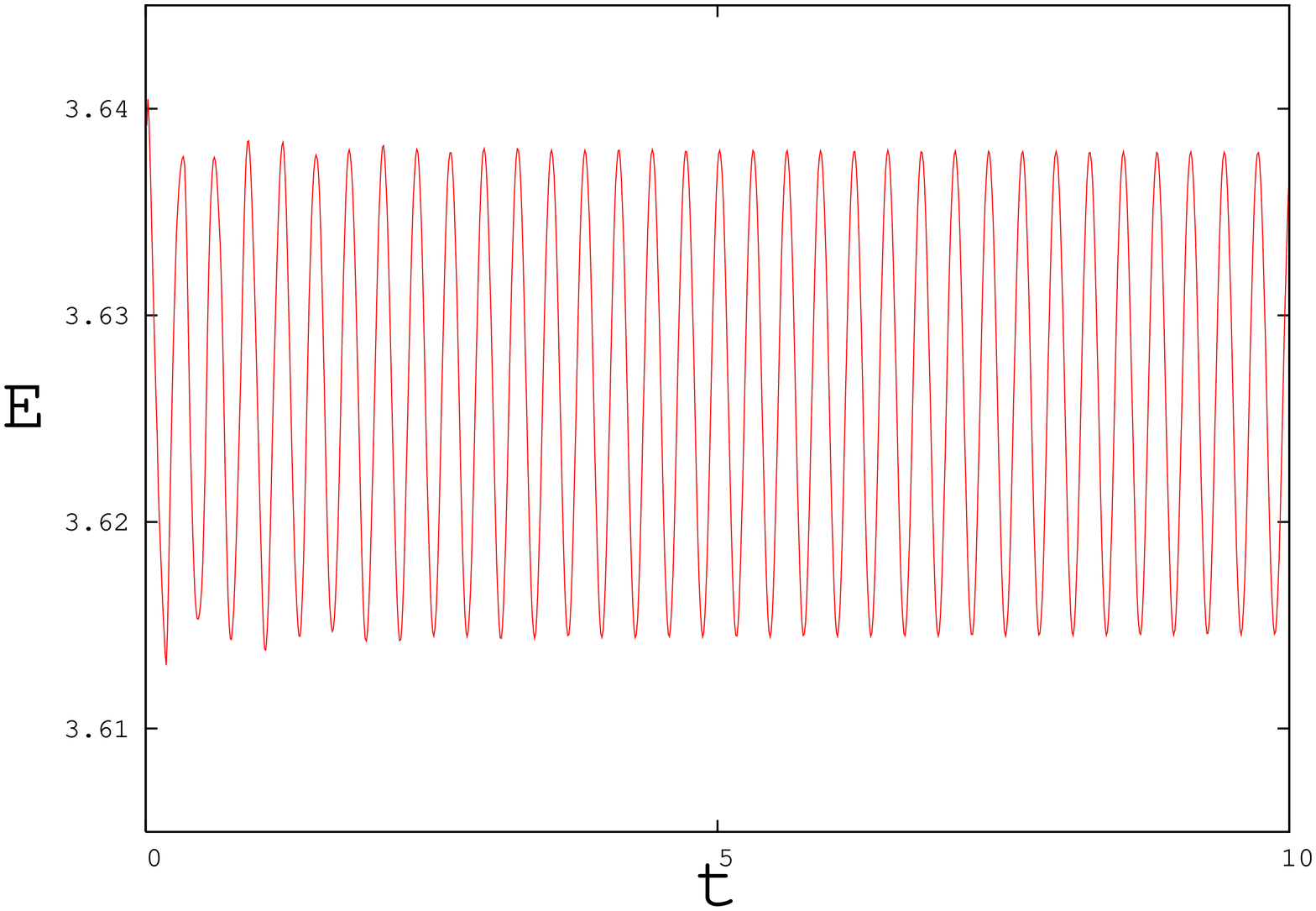,angle=270,width=3.0in}
\quad
\psfig{file=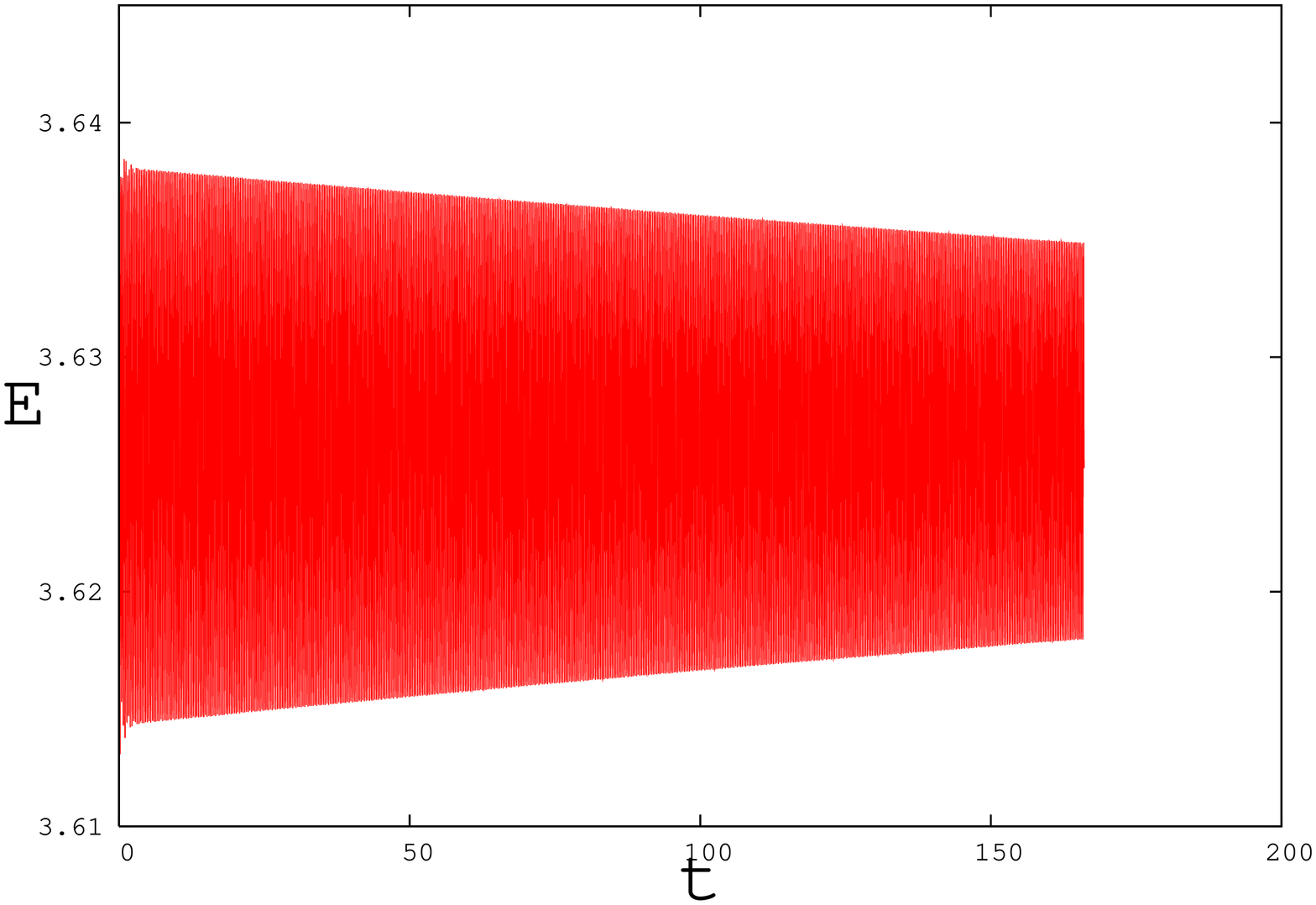,angle=270,width=3.0in}}
\caption{Temporal behaviour of the maximal amplitude $E_{max}$ of
the activator $E$ for model (\ref{karma}). The system length is just
below the Hopf bifurcation with $L=0.22$; the other parameters are as
in Figure~\ref{Fig-karma}. (a): The oscillations appear to be stable
over some time. (b): Same parameters as in (a) but longer integration
time. The apparent stability has to be accounted for by insufficiently
long integration times. The solution adjusts to the homogeneous
solution. Note the long time scales which contain hundreds of
oscillations.}
\label{Fig-karma2}
\end{figure}

\noindent
In the context of alternans the Hopf bifurcation had been described as
a supercritical bifurcation
\cite{Courtemanche,Karma93,Karma94,Courtemanche96} and not as we have
found here as a subcritical bifurcation (although at the same time
their occurrence had been related to wave breakup \cite{Karma94}). We
therefore revisit some of the previous numerical studies. In
\cite{Karma93} the following two-variable model was proposed
\begin{eqnarray}
\label{karma}
\epsilon \partial_t E&=&\epsilon^2\partial_{xx}E-E + \left[
A-\left(\frac{n}{n_B}\right)^M\right] \left( 1-\tanh(E-3)\right)
\frac{E^2}{2} \nonumber\\
\partial_t n&=&\theta(E-1)-n\; ,
\end{eqnarray}
as a model for action potential propagation in cardiac tissue. Here
$\theta(x)$ is the Heaviside step function. This model incorporates
essential features of electrophysiological cardiac models. For the
parameters $A=1.5415$, $\epsilon=0.009$, $M=30$ and $n_B=0.525$ a
supercritical Hopf bifurcation was reported upon diminishing the
system length $L$. We integrate this model using a pseudospectral
Crank-Nicolson method where the nonlinearity is treated with an
Adams-Bashforth scheme. We use a timestep of $dt=0.00001$ and $4096$
spatial grid points. A Hopf bifurcation occurs around $L=0.215$. To
approach the Hopf bifurcation we created a stable pulse for some large
system length, and subsequently diminished the system length $L$. In
Figure~\ref{Fig-karma} we show that for these parameters the
bifurcation is actually subcritical. The subcritical character has not
been recognized before - probably because of insufficiently short
integration times.
%
For system length $L$ just above the critical length the oscillations
can appear stable for a very long time (see Figure~\ref{Fig-karma2})
before they settle down to the homogeneous solution.
%
%

\begin{figure}
\centerline{
\psfig{file=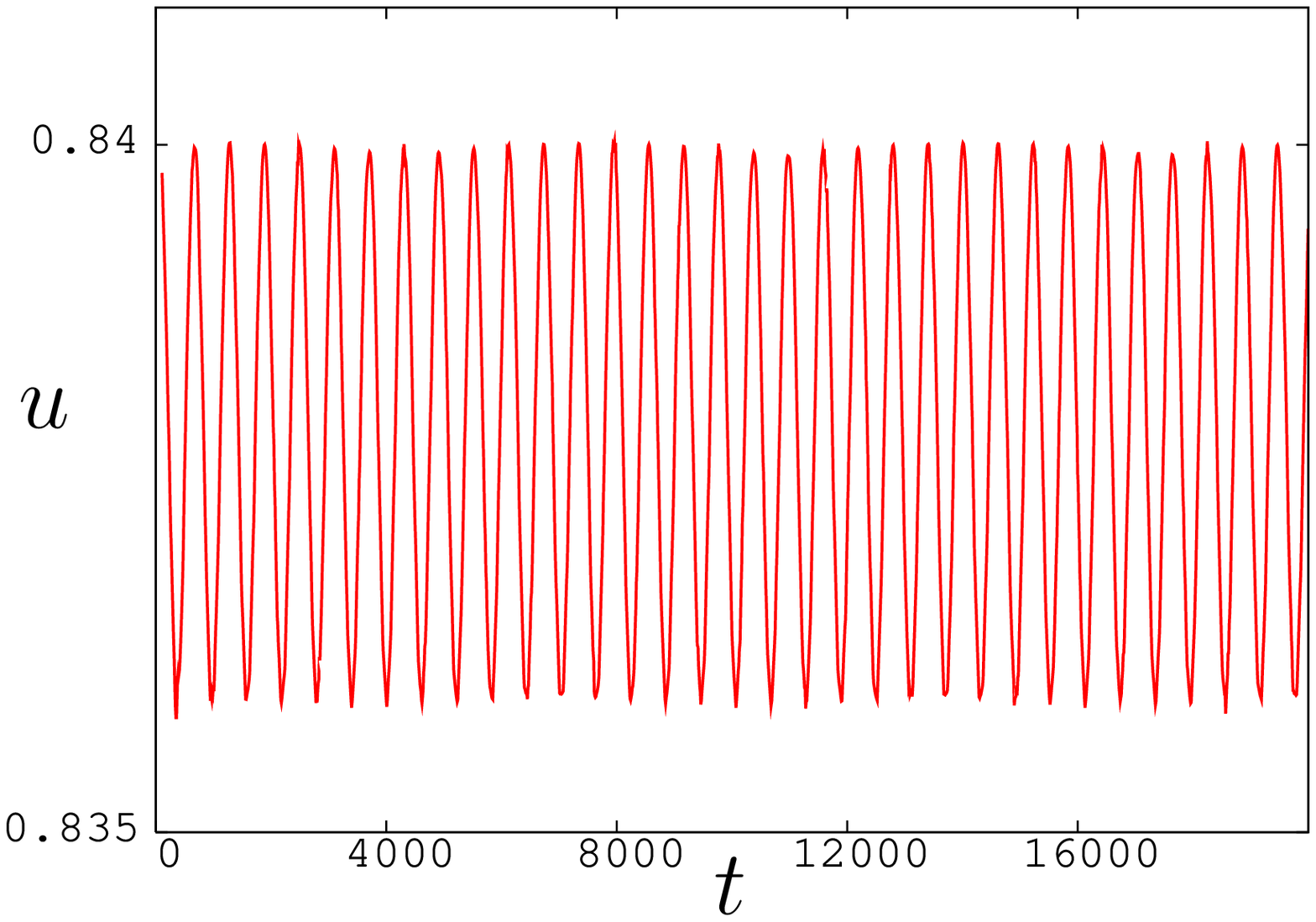,angle=0,width=3.0in}}
\caption{Temporal behaviour of the maximal amplitude $u_{max}$ of
the activator $u$ for model (\ref{barkley}) just above the subcritical
Hopf bifurcation. The parameters are $a=0.22$, $u_s=0.1$,
$\epsilon=0.03755$ and $L=246$. The oscillations appear stable for a
very long time but will eventually either damp out and attain a
constant non-zero value in the case, when $L$ is larger than the
critical $L_H$ at which the Hopf bifurcation occurs, or in the case
$L<L_H$ the pulse will collapse as depicted in Figure~\ref{Fig-karma}
confirming the subcritical character of the Hopf bifurcation.}
\label{Fig-barkley}
\end{figure}

Indeed, as already stated in our paper \cite{GottKram05}, the number
of oscillations may be rather large when the instability is weak. In
Figure~\ref{Fig-barkley} we show such a case for the maximal amplitude
of the activator $u$ for the modified Barkley model
\begin{eqnarray}
\label{barkley}
\partial_t u &=& D \partial_{xx}u + u(1-u)(u-u_s-v) \nonumber \\
\partial_t v &=&  \epsilon \ (u- a\  v)\  ,
\end{eqnarray}
which is a reparameterized version of a model introduced by Barkley
\cite{Barkley91}. It is clearly seen that the oscillations can appear
stable for a very long time and many oscillations (in this case more
than $500$ oscillations) which has lead scientists to the wrong
conclusion that the Hopf bifurcation is supercritical.\\

\begin{figure}
\centerline{
\psfig{file=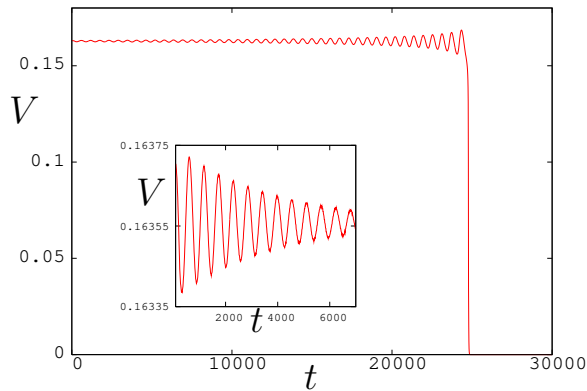,angle=0,width=3.0in}}
\caption{Temporal behaviour of the maximal amplitude $V_{max}$ of
the activator $V$ for model (\ref{EcheKarma1})-(\ref{EcheKarma3}). The
parameters are $\tau_0 = 150$, $\tau_a= 26$, $\tau_m = 60$, $\tau_p = 12$,
$V_c = 0.1$, $D= 0.00025$ and $\epsilon = 0.005$. The main figure is
obtained for $L=1.11$ which is slightly above the subcritical Hopf
bifurcation confirming the subcritical character of the Hopf
bifurcation. The inset is for $L=1.1175$ which is slightly below the
bifurcation point.}
\label{Fig-EcheKarma1}
\end{figure}

\begin{figure}
\centerline{
\psfig{file=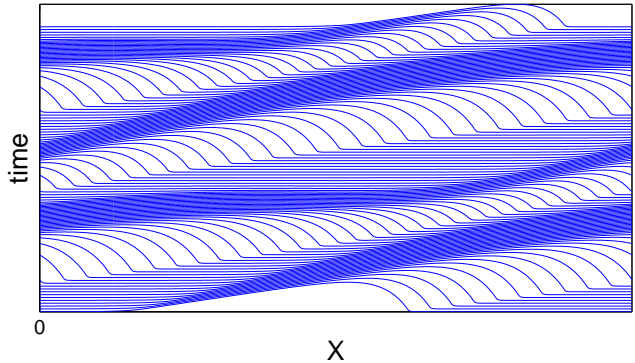,angle=0,width=4.0in}}
\caption{Space-time plot of stable oscillations occurring at
$\tau_a=6$ with $L=4.8$. The other parameters are as in
Fig.~\ref{Fig-EcheKarma1}. Stable oscillations are found for a range
of ring lengths $L$.}
\label{Fig-EcheKarma2}
\end{figure}

\begin{figure}
\centerline{
\psfig{file=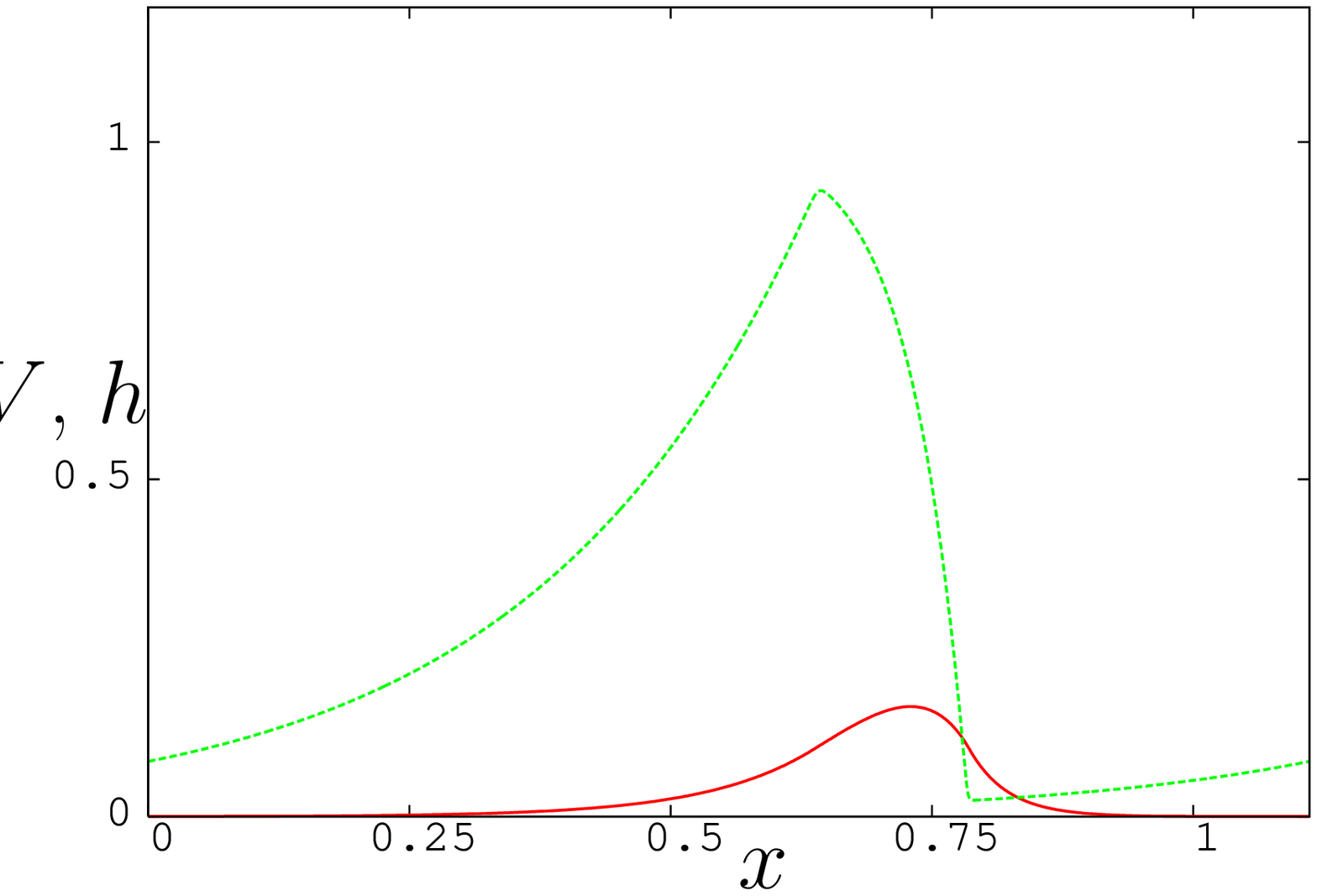,angle=0,width=2.5in}
\psfig{file=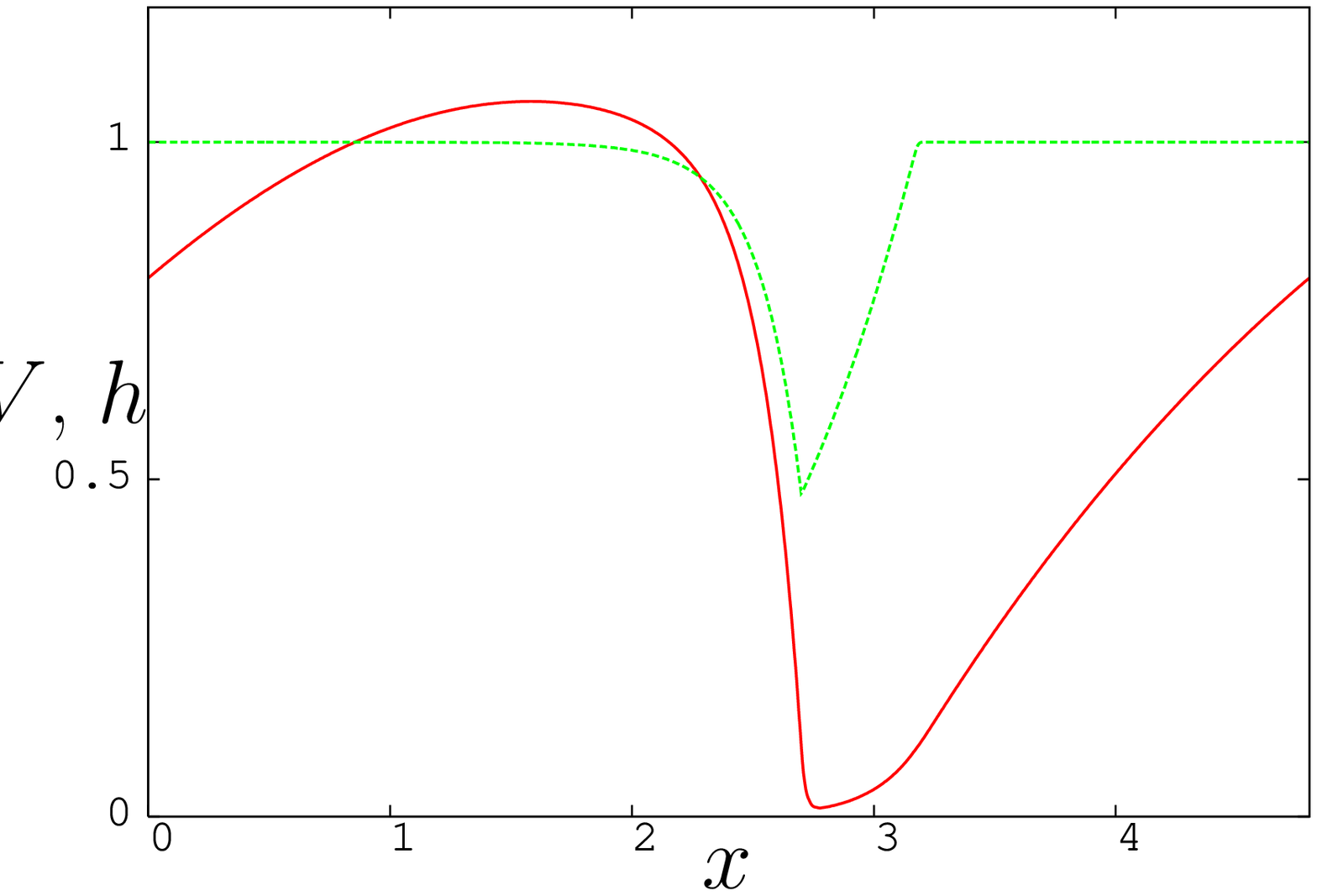,angle=0,width=2.5in}}
\caption{Plot of the activator $V$ (continuous line) and the inhibitor
$h$ (dashed line) for the system
(\ref{EcheKarma1})-(\ref{EcheKarma3}). We plot here $1-h$ rather than
$h$ to have the homogeneous rest state at $u=0$ and
$1-h=0$. Parameters are $\tau_0 = 150$, $\tau_m = 60$, $\tau_p = 12$,
$V_c = 0.1$, $D= 0.00025$ and $\epsilon = 0.005$. Left: The activator
runs into the exponentially decaying tail of the inhibitor which
decays towards the rest state $1-h=0$. This is similar to the
behaviour in Fig.~\ref{Fig-barkley2}. Parameters are $\tau_a=26$ with
$L=1.11$. This scenario is well described by the normal form. Right:
The activator does not interact with the exponentially decaying tail
corresponding to the rest state but rather with the metastable state
defined by ${\dot h}=0$. Parameters are $\tau_a=6$ with $L=4.8$. This
case cannot be captured by the normal form.}
\label{Fig-EcheKarma3}
\end{figure}

\begin{figure}
\centerline{
\psfig{file=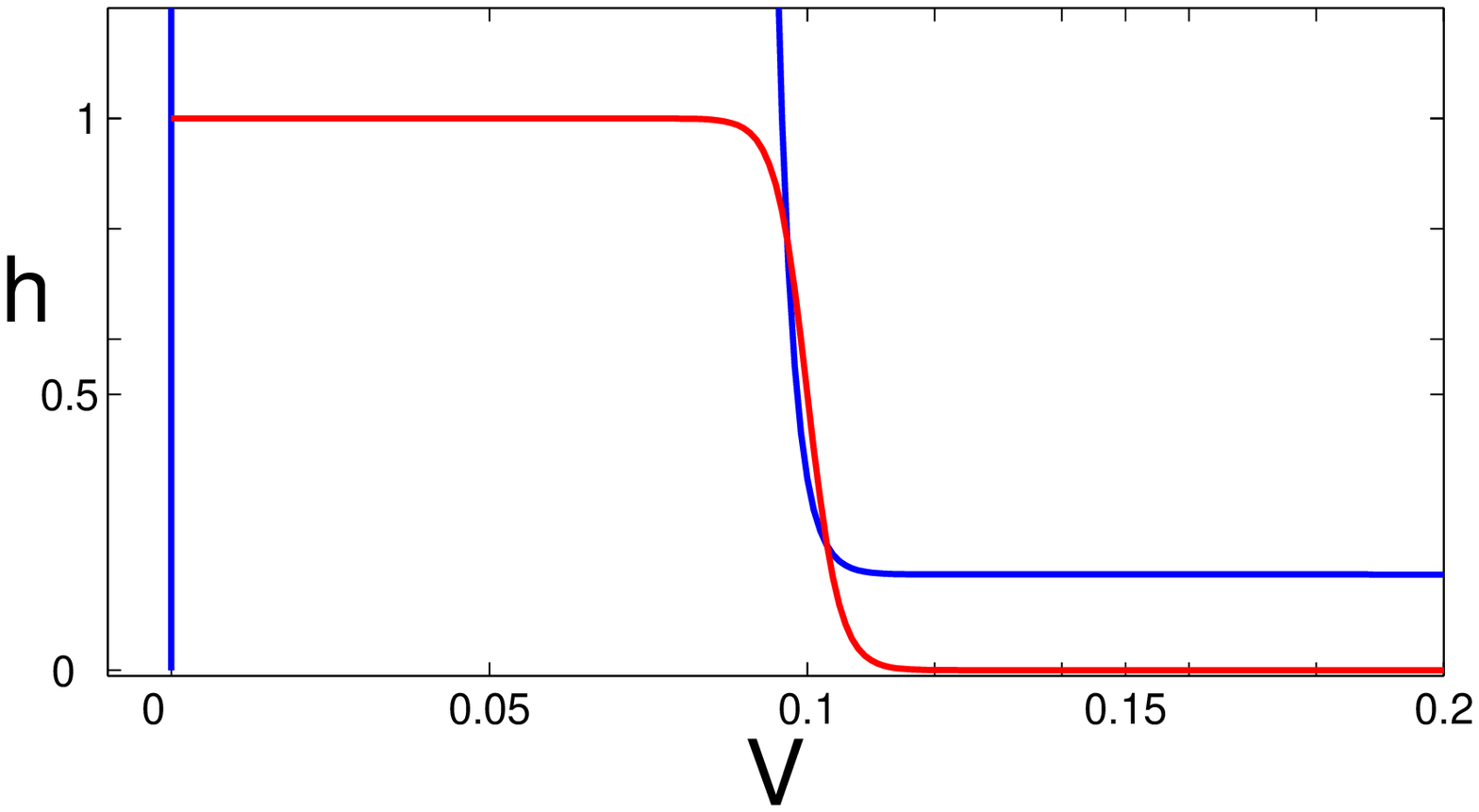,angle=0,width=2.5in,height=2.0in}
\psfig{file=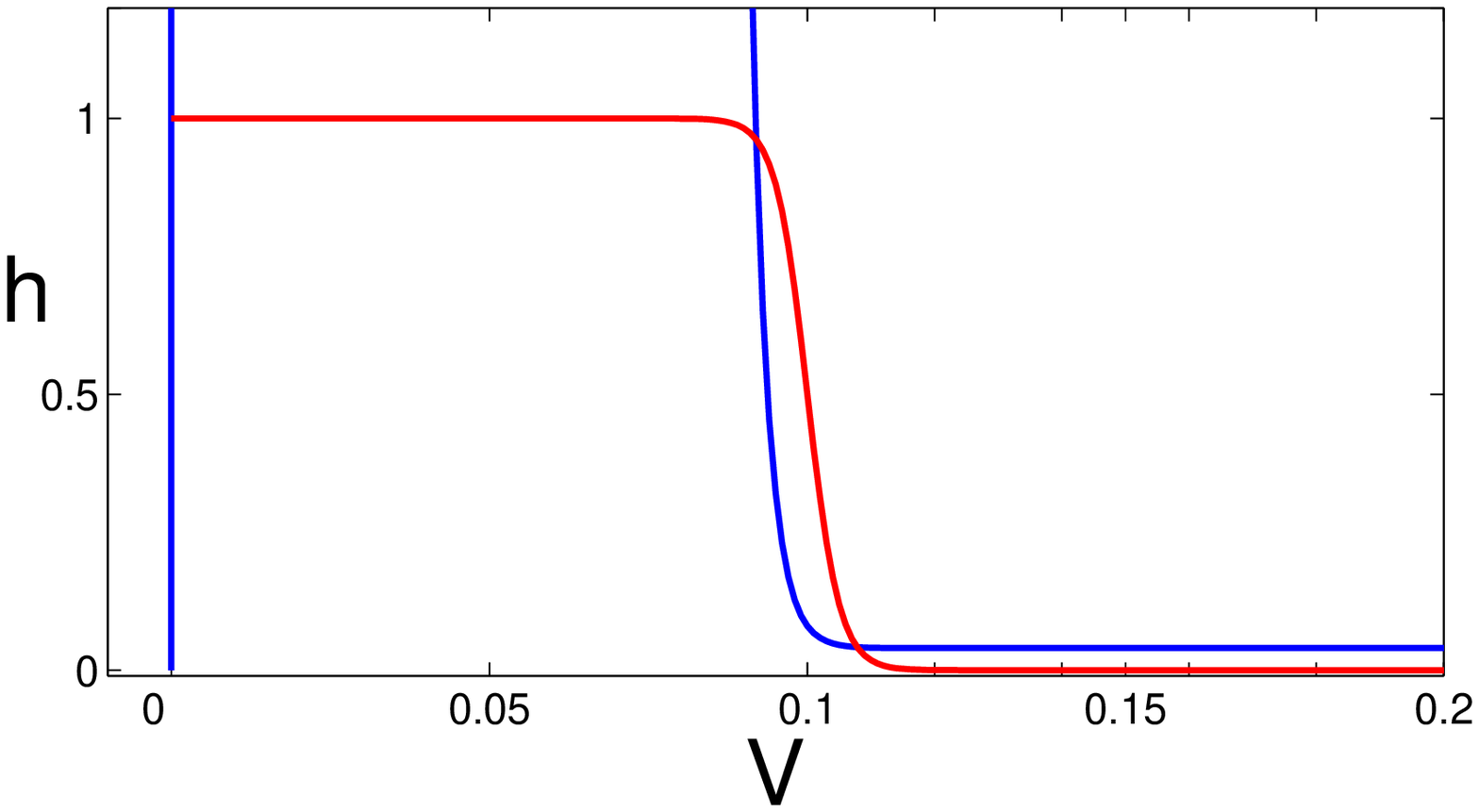,angle=0,width=2.5in,height=2.0in}
\psfig{file=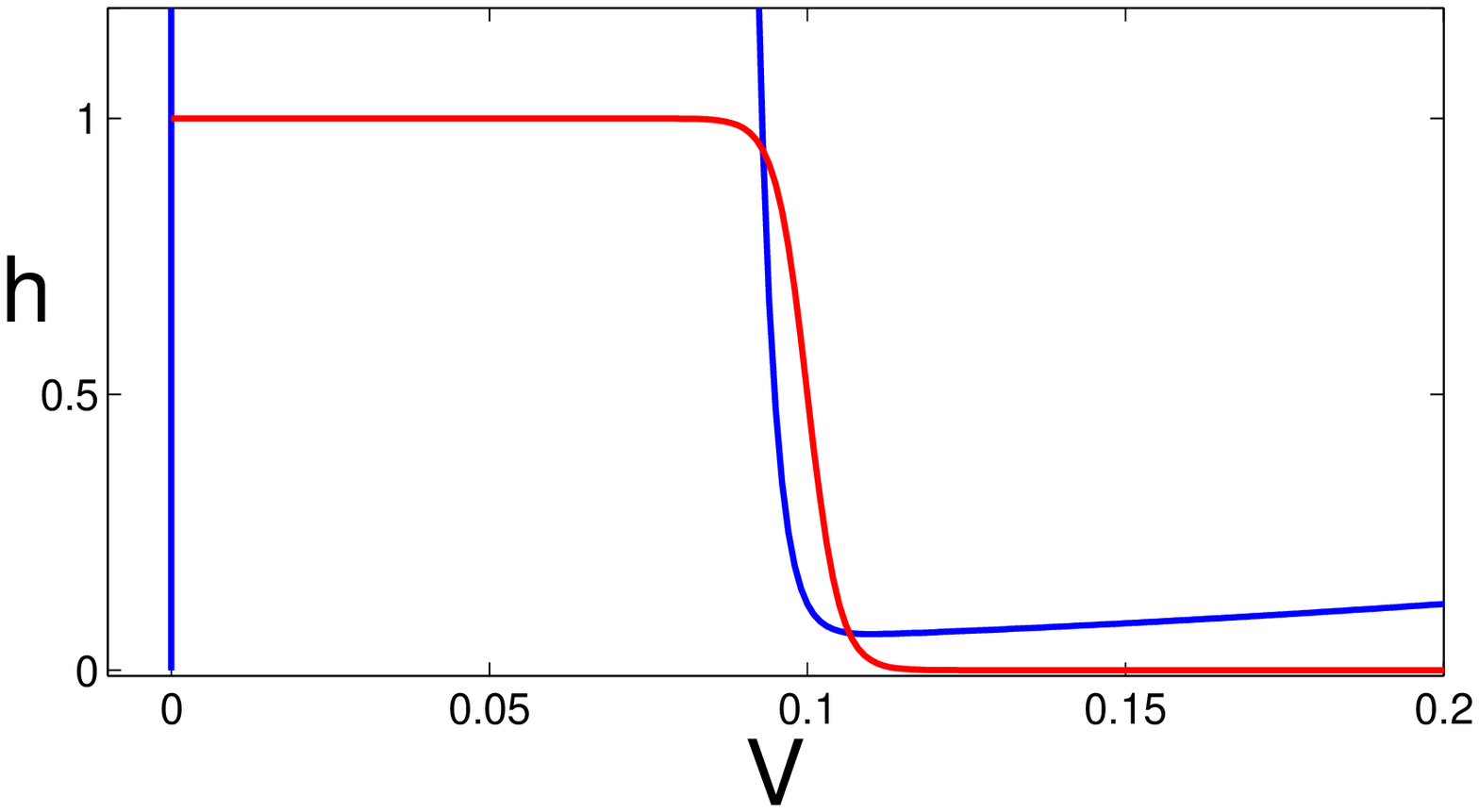,angle=0,width=2.5in,height=2.0in}
}
\caption{Nullclines for the system
(\ref{EcheKarma1})-(\ref{EcheKarma3}). The continuous lines denote the
${\dot V}=0$ nullclines and the dashed lines the ${\dot h}=0$
nullclines. Parameters are $\tau_0 = 150$, $\tau_m = 60$, $\tau_p =
12$, $V_c = 0.1$ and $\epsilon = 0.005$, and for all cases only one
stable fix point exists at $V=0$ and $h=1$. Left: The subcritical case
with $\tau_a=26$. Middle: The supercritical case with $\tau_a=6$. Note
the closeness of the nullclines for large $V$. Right: Nullclines for
the modification (\ref{EcheKarmacorr}) which breaks the near
degeneracy of the nullclines observed in the middle figure. Here
$\tau_a=6$ and $\tau_l=3$.}
\label{Fig-nullclines-echekarma}
\end{figure}

\noindent
The normal forms (\ref{NF}) or (\ref{NF2}) were derived for situations
in which the activator weakly interacts with the tail of the preceding
inhibitor which exponentially decays towards the homogeneous rest
state. Then one can describe the influence of the tail of the
preceding inhibitor as a perturbation to the generic saddle node of
the isolated pulse. The models discussed so far all fall into this
category. A different model was introduced by Echebarria and Karma in
\cite{Echebarria02} which as we will see below for certain
parameter regions does not fall into this class of model but supports
stable oscillations. Originally the model was studied for a paced
strand but recently has also been studied in a ring geometry
\cite{Echebarria06}. It has been argued in \cite{Echebarria06} that
the stability of the spatially extended pulse is determined by the
stability of a paced single cell. In the following we study
numerically the Hopf bifurcation for this model in a ring
geometry. This will illustrate the range of validity for our normal
form and the conclusions which may be drawn with respect to the
stability of cardiac alternans. The model consists of the standard
cable equation
\begin{eqnarray}
\label{EcheKarma1}
\partial_t V &=& D \partial_{xx}V - \frac{I_{{\rm ion}}}{C_m}\; ,
\end{eqnarray}
where $I_{\rm ion}$ models the membrane current and $C_m$ is the
capacity of the membrane. In \cite{Echebarria02} the following form
for the membrane current was proposed
\begin{eqnarray}
\label{EcheKarma2}
\frac{I_{{\rm ion}}}{C_m} = \frac{1}{\tau_0}\left(
S+(1-S)\frac{V}{V_c}\right)
-\frac{1}{\tau_a}hS
\; ,
\end{eqnarray}
with a switch function
\begin{eqnarray}
\label{EcheKarmaS}
S=\frac{1}{2}\left(1+\tanh(\frac{V-V_c}{\epsilon})\right)
\; .
\end{eqnarray}
The gate variable $h$ evolves according to
\begin{eqnarray}
\label{EcheKarma3}
\frac{dh}{dt}=\frac{1-S-h}{\tau_m(1-S)+\tau_pS}
\; .
\end{eqnarray}
The stable homogeneous rest state is at $V=0$ and $h=1$; however for
small $\tau_a$ a second stable focus may arise. For details on the
physiological interpretations of the model the reader is referred to
\cite{Echebarria02,Echebarria06}. For the numerical integration we use
again a semi-implicit pseudospectral Crank-Nicolson method where the
nonlinearity is treated with an Adams-Bashforth scheme. We use a
timestep of $dt=0.01$ and $1024$ spatial grid points. In
Fig.~\ref{Fig-EcheKarma1} we show an example for a subcritical Hopf
bifurcation in this model consistent with our theory. However, for
sufficiently small $\tau_a$ a supercritical Hopf bifurcation arises
upon decreasing the ring length $L$. In Fig.~\ref{Fig-EcheKarma2} we
present a space-time plot for such a situation of stable oscillations.
Whereas the subcritical case is consistent with our theory we now have
to understand why for small $\tau_a$ stable oscillations occur. In
order to do so it is helpful to look at the spatial profiles of the
activator and the inhibitor close to the Hopf bifurcation which are
presented in Fig.~\ref{Fig-EcheKarma3}. In the left figure we see the
activator $V$ and the inhibitor $1-h$ for the case of a subcritical
Hopf bifurcation as seen in Fig.~\ref{Fig-EcheKarma1}. The figure is
similar to Fig.~\ref{Fig-barkley2} for the modified Barkley model. The
activator weakly interacts with the exponentially decaying tail of the
inhibitor it created during its previous revolution. In this parameter
region our normal form is valid and correctly predicts a subcritical
bifurcation. In the right figure of Fig.~\ref{Fig-EcheKarma3} the
situation is depicted for the supercritical case seen in
Fig.~\ref{Fig-EcheKarma2}. Here the situation is very different. The
inhibitor does not approach the homogeneous rest state $1-h=0$ but
rather develops a metastable $1-h=1$ plateau. This has two
consequences; firstly, the solution is driven away from the homoclinic
pulse solution around which the normal form is built, and secondly the
interaction is not weak anymore.
The reason for this different behaviour can be understood by looking
at the nullclines of the homogeneous problem of
(\ref{EcheKarma1})-(\ref{EcheKarma3}), i.e. setting $\partial_x=0$. In
Fig.~\ref{Fig-nullclines-echekarma} we show the nullclines for the two
cases $\tau_a=6$ (supercritical) and $\tau_a=26$ (subcritical). Note
that for $\tau_a=6$ the only stable fix point is at $V=0$ and
$h=1$. The difference is that in the supercritical case the ${\dot
h}=0$ nullcline and the ${\dot V}=0$ nullcline are very close to each
other. This forces the trajectory to spend a long time on the ${\dot
h}=0$-nullcline near $h=0$ (as seen in the plateau part of the spatial
profile of $1-h=1$ in Fig.~\ref{Fig-EcheKarma2}). We call this state a
metastable state. For decreasing ring length it dominates the profile
of the inhibitor and does not allow the inhibitor to come close to the
rest state $h=1$. Therefore our normal form, which is formulated
around the saddle-node of the pulse, breaks down. The solution is not
close to the travelling pulse in phase space anymore and our local
analysis around the saddle-node of the travelling wave cannot work
anymore. However, we note that the system
(\ref{EcheKarma1})-(\ref{EcheKarma3}) is rather unusual with the two
nullclines being parallel to each other with the possibility of a
metastable state, resulting in rather particular dynamical
behaviour.\\

We may modify the model (\ref{EcheKarma1})-(\ref{EcheKarma3}) to break
the degenerate situation in which the two nullclines of the inhibitor
and activator run parallel to each other and subsequently may get too
close to each other for certain parameters. We can destroy the
existence of the metastable state for finite ring length by allowing
the nullcline of the activator to bend away from the nullcline of the
inhibitor if we, for example, consider the following modification of
the membrane current
\begin{eqnarray}
\label{EcheKarmacorr}
\frac{I_{{\rm ion}}}{C_m} = \frac{1}{\tau_0}\left(
S+(1-S)\frac{V}{V_c}\right) 
+ \frac{1}{\tau_l}V^2
-\frac{1}{\tau_a}hS
\; ,
\end{eqnarray}
with some sufficiently small $\tau_l$. Then we are again in the
situation where the rest state $V=0$ and $h=1$ dominates the dynamics
upon decreasing the ring length $L$. The nullclines are shown in
Fig.~\ref{Fig-nullclines-echekarma}. We confirmed that for $\tau_l=3$
the Hopf bifurcation is indeed subcritical, consistent with our
theoretical result. We note that the actual value of $\tau_l$ is not
important for the existence of subcritical bifurcation but rather that
a sufficiently small $\tau_l$ breaks the geometric structure of the
degenerate nullclines and allows the activator nullcline to bend away
from the inhibitor nullcline. The model
(\ref{EcheKarma1})-(\ref{EcheKarma3}) illustrates for which class of
excitable media our normal form is applicable and for which systems we
may draw conclusions on the stability of dynamical alternans in a
ring.\\


\medskip

{\underbar{\bf Acknowledgements }} I would like to thank Sebastian
Hermann for helping with the DDE-BIFTOOL software, and Martin
Wechselberger for fruitful discussions. I gratefully acknowledge
support by the Australian Research Council, DP0452147 and DP0667065.


\newpage


\end{document}